\documentclass[usenatbib,useAMS,usegraphicx]{mn2e}

\input{journal.cls}

\newcommand{\bm}[1]{\mbox{\boldmath$ #1 $}}
\newcommand{\lskip}{\vskip \baselineskip}
\newcommand{\nskip}{\lskip \noindent}
\newcommand{\halfskip}{\vskip 0.5\baselineskip}
\newcommand{\be}{\begin{equation}}
\newcommand{\ee}{\end{equation}}
\newcommand{\SWa}{\mbox{\rm SW1}}
\newcommand{\SWb}{\mbox{\rm SW2}}

\newcommand{\equref}[1]{(\ref{#1})}
\newcommand{\RNum}[1]{\uppercase\expandafter{\romannumeral #1\relax}}
\newcommand{\NTracer}{\mbox{$\mathcal{N}$-Tracer} }
\newcommand{\NCooling}{\mbox{$\mathcal{N}$-Cooling} }

\newcommand{\epsprime}{\mbox{$\epsilon'$}}

\newcommand{\rhoprime}{\mbox{$\rho'$}}
\newcommand{\nprime}{\mbox{$n'$}}

\newcommand{\grad}{\mbox {\boldmath $\nabla$}}
\newcommand{\bdot}{\mbox{$\bm{\: \cdot \:}$}}

\usepackage{hyperref}
\usepackage[table]{xcolor}
\usepackage{graphicx,epsfig}
\usepackage{verbatim}
\usepackage{syntonly}
\usepackage{txfonts}
\usepackage{caption}
\DeclareCaptionLabelFormat{continued}{#1~#2~(cont.)}
\title[Rel. AGN jets III. Synchrotron emission from DDRGs]%
{Relativistic AGN jets III. Synthesis of synchrotron emission %
from Double-Double Radio Galaxies}

\author[S. Walg et al.]{
S. ~Walg, $^{1,2}$\thanks{email: \url{sanderwalg@gmail.com}}
A. ~Achterberg, $^1$
S. ~Markoff, $^2$
R. ~Keppens, $^3$
O. ~Porth $^2$
\\
$^1$Department of Astrophysics/IMAPP, Radboud University, P.O. Box 9010,
6500 GL Nijmegen, The Netherlands.
\\
$^2$Astronomical Institute "Anton Pannekoek," University of Amsterdam,
                                   Science Park 904, 1098 XH Amsterdam, The Netherlands
\\
$^3$Centre for mathematical Plasma Astrophysics, Department of Mathematics, KU Leuven,
                                   Celestijnenlaan 200B, 3001 Heverlee, Belgium
}

\date{\today $\;\;$ (accepted 23 July 2020, DOI: 10.1093/mnras/staa2195)}

\pagerange{\pageref{firstpage}--\pageref{lastpage}}
\volume{0000}
\pubyear{2013}

\begin{document}

\maketitle

\label{firstpage}


\begin{abstract}
The class of Double-Double Radio Galaxies (DDRGs) relates to episodic jet outbursts.
How various regions and components add to the total intensity in radio images is less
well known. In this paper we synthesize synchrotron images for DDRGs based on special
relativistic hydrodynamic simulations, making advanced approximations for the magnetic
fields. We study the synchrotron images for:
Three different radial jet profiles;
Ordered, entangled or mixed magnetic fields;
Spectral ageing from synchrotron cooling;
The contribution from different jet components;
The viewing angle and Doppler (de-)boosting;
The various epochs of the evolution of the DDRG.
To link our results to observational data, we adopt to J1835+6204 as a reference source.
In all cases the synthesized synchrotron images show two clear pairs of hotspots, in
the inner and outer lobes. The best resemblance is obtained for the piecewise isochoric
jet model, for a viewing angle of approximately $\vartheta \sim -71^{\circ}$,
i.e. inclined with the lower jet towards the observer, with predominantly entangled
($\gtrsim 70$ per cent of the magnetic pressure)  in turbulent, rather than ordered
fields. The effects of spectral ageing become significant when the ratio of observation
frequencies and cut-off frequency $\nu_{\rm obs}/\nu_{\infty,0} \gtrsim 10^{-3}$,
corresponding to $\sim 3 \cdot 10^2$ MHz. For viewing angles
$\vartheta \lesssim -30^{\circ}$, a DDRG morphology can no longer be recognized. The
second jets must be injected within~$\lesssim$~4~per~cent of the lifetime of the first
jets for a DDRG structure to emerge, which is relevant for Active Galactic Nuclei
feedback constraints.
\end{abstract}



\section{Introduction}
\label{sec:Introduction} 
Active Galactic Nuclei (AGN) jets that are created by the central engine of a
powerful radio galaxy are capable of growing up to megaparsec scales
\citep{Dabhade2017}. They can be active for tens or hundreds of
megayears~\citep{Bird2008,Turner2015}. Some radio galaxies are thought to have gone
through an episodic outburst cycle, where the central engine has been turned off for
some time. If this time of intermittency is less than the time it takes for a jet plasma
element to travel from the central engine to the jet-head, there will be a phase only
seen in radio images where two distinct jets and counter jets will be visible, as well
as their corresponding hotspots, forming a double-double radio galaxy (DDRG).
\lskip

The class of double-double radio galaxies has been discovered a relatively short time ago
(\citealt{Schoenmakers2000a},\citealt{Schoenmakers2000b}; \citealt*{Kaiser2000}).
It is characterized by having two aligned distinct pairs of radio lobes originating from
the same central engine of the AGN. More than 20 DDRGs have been clearly identified so
far (see for example
\citealt{Saikia2009},
\citealt{Nandi2019},
\citealt{Kuzmicz2017} or
\citealt{Mahatma2019}),
but with the increasing sensitivities and radio bandwidth in novel radio telescopes such
as LOFAR, the number of detected radio loud AGN or LOFAR DDRGs has increased significantly
(see for example the GLEAM survey, \citealt{Callingham2017}; or the LOFAR Two-metre Sky
Survey (LoTSS), see
\citealt{Shimwell2019}, \citealt{Hardcastle2019}; \citealt{Mahatma2019}) so that the
number of detectable DDRGs might also grow significantly \citep{Orru2015}. As the
synchrotron emission fades rapidly, the higher sensitivity and low frequency coverage as
achieved by LOFAR is crucial in that aspect. The typical morphology of these sources, as
well as studies on the spectral age of different regions, suggest that DDRGs are the
result of episodic jet
activity~\citep{Mahatma2019,Schoenmakers2000a,Saikia2009,Marecki2016,Nandi2017,Nandi2019}.
The jet size can be extremely large, while the associated $\sim$ Myr time scales for
source evolution excludes us to witness the episodic scenario within a single, individual
source.

The radio emission observed in DDRGs is mainly (non-thermal) synchrotron radiation. It is
produced by relativistic electrons that are spiralling along magnetic field lines which
are carried along by the plasma. What we see is the sum (superposition) of all
synchrotron-radiating particles. However, for jets with a radial structure and in the
case of episodic activity, different jet components might have very different
intensities, leading to different brightness variations in synchrotron radiation.
Brightness variations in radio images and the frequency at which the source is observed
can strongly influence the general observed morphology of these sources. This can be a
complicating factor when comparing an observed DDRG to a synthesized map of a numerical
simulation. To better understand the processes that lead to the observed DDRGs, a study
on the synchrotron radiation coming from various jet components is essential.
\lskip

In the past few decades, a number of numerical studies on episodic jet activity have
been performed (see e.g.
\citealt{Wilson1984};
\citealt{Clarke1991};
\citealt{Chon2012};
\citealt*{Mendygral2012};
\citealt{Refaelovich2012};
\citealt*{Stepanovs2014};
\citealt{Walg2014};
\citealt{Seong-Jae2016}).
In \citet{Walg2014}, a 2.5$D$ (i.e. accounting for all three flow components but assuming
axisymmetric evolution) cylindrical symmetric special relativistic hydrodynamical (SRHD)
study was performed on episodic activity of relativistic AGN jets. In that work, the
density, pressure and jet velocity were chosen such that they correspond well to those
inferred from observations of a typical DDRG (for example \citealt{Konar2013a}). The jets
discussed in \cite{Walg2013} (hereafter $\SWa$) and \cite{Walg2014} (hereafter $\SWb$)
have radial structure, consisting of a high-Lorentz factor jet spine and slower jet
sheath. This is inspired from observations, showing evidence for spine-sheath structured
jets~\cite[see e.g.][]{Laing2014}. The main emphasis in $\SWb$ was on the integrity of
this structure as the jet propagates, and how the different jet components mix.

  \subsection*{Main focus of this research}

Synchrotron emissivity is determined by the number density of \mbox{relativistic}
particles, the frequency of the photons and the strength of the magnetic field
component perpendicular to the particle's velocity. Bulk motion of the source
material, and in particular motion along the line-of-sight, influences the
observed intensity through the Doppler effect. This will be discussed in more
detail in section \ref{sec:TheoreticalBackground}. In order to
create a synthesized synchrotron map, one requires a full special relativistic
magnetohydrodynamic (SRMHD) treatment with additional dynamical tracing of a
relativistic particle population. Presently, there are a great number of
relativistic magneto-hydrodynamics (MHD) codes. A recent study compares a significant
number of (general relativistic) magneto-hydrodynamics (GRMHD) codes including
{\small Athena++,
BHAC,
Cosmos++,
ECHO,
H-AMR,
iharm3D,
HARM-Noble,
IllinoisGRMHD,
Koral} 
\citep{Porth2019},
but in this study GR is not needed when studying the larger scales in the jets.
Despite the benefits of having detailed data of the evolution of the magnetic
fields, SRMHD simulations are computationally much more expensive than pure
special relativistic hydrodynamic (SRHD) simulations.

\lskip
Close to the AGN, magnetic fields are dynamically important and may be linked to
the mechanism that collimates the jet flow (e.g.
\citealt*{Begelman1984};
\citealt{Bromberg2011};
\citealt*{Blandford2019};
\citealt{Chatterjee2019}
). At large (Mpc)
scales, when the jet has reached its terminal velocity, the fields are no longer
dynamically important as the jet flow is expected to reach strongly
supersonic/super-Alfv\'enic speeds. Moreover when the jets have switched off, the plasma
will mostly evolve adiabatically. In that case, we make the often used assumption that
there is equipartition between the pressure stored in the magnetic field and the gas
pressure from the relativistic particle population. From our SRHD simulation, this is the
simplest way to handle the unknown magnetic field strength quantitatively, and
equipartition fields are close to the state in which the total energy of the plasma is
minimized. We will distinguish between two underlying magnetic field topologies, namely
ordered versus turbulent field, which together build up the deduced magnetic pressure.
In SRMHD simulations, one would need excessive resolutions to properly account for
turbulent magnetic fields. From our grid-adaptive SRHD runs, we get extended turbulent
mixing regions as discussed in $\SWa$ and $\SWb$. Although some
studies~\citep{Croston2005,Croston2018,Mahatma2020} show evidence for field strengths in
the lobes that are lower than equipartition estimates, we find it reasonable to assume
that the gas pressure from the relativistic particles is directly proportional to the
mass fraction of the different jet components. In the numerical simulations in this
paper, we have used additional tracers, to identify the various jet components at all
times, allowing to determine mass fractions of jet material in different jet episodes or
different layers (spine versus sheath). This allows us to infer the magnetic field
strength directly from the bulk gas pressure and tracer values, without the need for
dynamically evolving the magnetic fields.

The synchrotron emission from AGN jets can usually be represented by a power law
in frequency $\nu$,
\be
	S_{\nu} \propto \nu^{- \alpha} \; ,
\ee
over a wide range of frequencies, where $\alpha$ is the spectral index. This is
the emission expected from a population of relativistic electrons with an energy
distribution
$N(\epsilon) \: {\rm d}\epsilon \propto
\epsilon^{- (2 \alpha +1)} \: {\rm d} \epsilon$
(e.g. \citealt{Longair2011}, Ch. 8, or \citealt{RybickiLightman1986}, Ch. 6). Here
$\epsilon$ is the electron energy. The power-law continues up to some cut-off frequency
$\nu_{\rm \infty}$, where a break in the spectrum occurs either due to synchrotron
losses of the relativistic electrons or the fact that the acceleration mechanism has
a limiting (maximum) energy.
\lskip

When a power law distribution is assumed for the population of relativistic particles at
the jet inlet, the evolution of this population can be calculated in two ways:
i) either it can be inferred from the bulk mass density and the mass fraction of jet
material;
ii) it can be deduced from advecting a quantity proportional to the cut-off energy
$\epsilon_{\infty}$ (see for example \citealt{CamusThesis}, Ch. 4;
or \citealt{Camus2009}). Both methods will be discussed in this paper.
We utilize the tracers for various jet components to full extent. This allows us
to study the influence of the various jet components on the brightness variations
within a synchrotron map. Moreover, we will impose various magnetic field
configurations on our numerical simulations and study the influence on the overall
morphology. Finally, we utilize the jet tracers to give an estimation of how
large-scale ordered fields transform into entangled fields, so a mixed magnetic
field configuration can be used. In this work we focus on synthesizing synchrotron
maps from the same simulations that were presented in $\SWb$. In $\SWb$, we emphasized
that subtly different dynamics and mixing properties arise when jets that differ in their
radial stratification interact. This will undoubtedly reflect in synthetic synchrotron
views, and this will become evident in this paper. The synchrotron views are expected to
be sensitive to the deduced magnetic field configurations, an aspect to be investigated
here as well. Moreover, radiative losses (synchrotron ageing) can be a decisive factor
in the overall radio appearance. Therefore, we will:

\begin{itemize}
\item compare the emission from three different jet models:
a homogeneous jet;
a piecewise isochoric jet;
and an isothermal jet;
\item compare the emission for various different imposed magnetic field
configurations;
\item study the effect of dimming due to synchrotron ageing of the first jet on
the radio morphology;
\item compare the emission coming from the initial jet and the restarted jet
separately;
\item consider different viewing angles for the DDRG phase, taking into account
the effect of the Doppler factor;
\item consider the emission from the source at various phases of the outburst
cycle.
\end{itemize}

As a reference DDRG source, we will consider J1835+6204, (observed at
4860.1~MHz with the VLA in the B or D configuration, with beam sizes 1.14'' and
1.71''), for which various observational results exist in the
literature~\citep{Konar2012,Konar2013b}.
In~\cite{Konar2013b}, the duration of the quiescent phase was estimated to be
$\sim$ 1 Myr, as short as $4.5$ per cent of the previous active phase, while the
age of the source is estimated to be $\sim$ 22~Myr. Estimates of the speed of the
jet spine, from a sample of radio galaxies including J1835+6204, were made
in~\cite{Konar2013a}, inferring Lorentz factors even above 10, to explain the
observations. The spectral index for this source is $\alpha \sim 0.818$.
Our models for DDRG jets discussed in $\SWb$ adopted parameters inspired from
observations, with e.g. the first jet injection lasting for 15.3 Myr, a (smoothly
varying) switching-off of the central engine between 15.3 and up to 16 Myr (see
Eqns.~(3)-(4) and Fig.~2 in $\SWb$), and a subsequent jet injection for another
6.8 Myr. The Lorentz factor is adopted as 3.11 for the homogeneous jet, while the
spine--sheath models (isochoric or isothermal jets) have a central spine with a
Lorentz factor 6. All three jet models derive parameters from a typical kinetic
jet luminosity of a few $\times 10^{46} \,{\mathrm{erg}} \,{\mathrm{s}}^{-1}$.
Details of the jet models are described in $\SWa$, where it was shown how
different internal jet structures can lead to dramatic differences in maintaining
jet integrity, and on their overall propagation speeds. In $\SWb$, the same jet
stratification models were then used to discuss their interaction in a DDRG
scenario, and here we follow up with their synthetic radio views.

\section{Theoretical background}
\label{sec:TheoreticalBackground}

We assume that the bulk flow of jet material contains a population of
relativistic particles. A recent study suggests that the composition of
relativistic jets in FR-II radio galaxies consists of electrons, positrons
and protons \citep*{Kawakatu2016}. In our simulations, we do not consider low-power
(FR-I class) jets, which may entrain significantly more material while propagating
through the host galaxy and the Intergalactic medium \citep{Laing2014}.
Such jets are more vulnerable to instabilities at the jet-cocoon and cocoon-intergalactic
medium interfaces, with resultant turbulent mixing. In our FR-II scenarios, we model all
entrainment with the surrounding medium in the simulations, except for any entrainment
aspects close to the central engine, which we exclude. We treat synchrotron emission in
the optically thin limit.
The total intensity of synchrotron radiation, $I(X,Y)$, at sky position $(X,Y)$
and frequency $\nu$ is completely determined by the synchrotron emissivity
coming from the plasma of non-thermal electrons. Hereafter, we shall denote
$\bm{n}$ as a unit vector along the line of sight pointing at the observer,
measured in the observers frame. The intensity measured at the sky position
$(X,Y)$ is calculated by integrating the emissivity $j_{\nu}$ along the line
of sight \mbox{$Z\bm{n}$}, perpendicular to the sky coordinate plane. It equals:

\be
I_{\nu}=\int \: {\rm d}Z \: j_{\nu}(\bm{r}\:,\:t) \; ,
\ee
where $j_{\nu}(\bm{r}\:,\:t)$ is the synchrotron emissivity at time $t$ and
position \mbox{$\bm{r}\: =\: (X,\:Y,\:Z)$} and frequency $\nu$, all as measured in
the observer frame. The effects of synchrotron self-absorption at low frequencies
are neglected. In the rest-frame of the plasma (where we denote
observables with an apostrophe \rq{} ), the synchrotron emissivity is given by:

\be
j_{\nu'}' \propto \mathcal{N} \: (B_{\perp}')^{\alpha + 1} \: (\nu')^{-\alpha} \;
\label{eq:syncemissrest}
\ee
(for instance: see \citealt{Longair2011}). Here, $\mathcal{N}$ is proportional to
the density of the relativistic particles (see appendix
\ref{sec:RelativisticParticles}), $B_{\perp}'$ is the magnetic field strength
perpendicular to the particle's velocity in the rest-frame of the plasma (see
appendix \ref{sec:MagneticField}), $\nu'$ is the photon frequency and $\alpha$,
as before, the spectral index of the plasma.

In the observer frame, however, Doppler \mbox{(de-)boosting} needs to be taken
into account. The synchrotron emissivity and the photon frequency are Doppler
shifted according to (\citealt{Begelman1984} and \citealt{RybickiLightman1986}):

\be
j_{\nu} = \mathcal{D}^2 j_{\nu'}'
\ee
and
\be
\nu     = \mathcal{D} \nu' \; ,
\ee
so that the emissivity observed at frequency $\nu$ is given by:

\be
j_{\nu} \propto \mathcal{D}^{2+\alpha} \mathcal{N} \:
(B_{\perp}')^{\alpha + 1} \:(\nu)^{-\alpha} \; .
\label{eq:jnuGeneral}
\ee
The Doppler factor $\mathcal{D}$ (for a smooth continuous jet) is given by:

\be
\mathcal{D} = \frac{1}{\gamma (1 - \beta \cos{(\psi}))} \;
= \frac{1}{\gamma (1 - \bm{\beta} \bdot \bm{n})} \; ,
\label{eq:DopplerFactor}
\ee
with \mbox{$\gamma({\bm r}\:,\:t)\:=\:1/\sqrt{1-\bm{\beta}\bdot\bm{\beta}}$} the
bulk Lorentz factor,
\mbox{$\bm{\beta}({\bm r}\:,\:t)\: =\: \bm{V}({\bm r}\:,\:t)/c$} the bulk
3-velocity of the plasma in units of $c$, and
\mbox{$\psi \: = \angle (\bm{n},\:\bm{\beta})$} the angle between the line of
sight and the bulk velocity of the plasma. 
\lskip

Based on a number of assumptions, we approximate various magnetic field
configurations from the hydrodynamic quantities. Near the central engine
of an AGN, magnetic fields are believed to have a helical structure (see for
example
\citealt{Blandford1977};
\citealt{Blandford1982};
\citealt{Keppens2008};
\citealt{McKinney2009};
\citealt{Tchekhovskoy2011};
\citealt{Laing2014};
\citealt{Zamaninasab2014};
\citealt{Gabuzda2015};
\citealt{Prior2019}).
At large distances from the central engine the large-scale magnetic
field structures are much more difficult to determine: the way that the
large-scale structures evolve is strongly dependent on jet dynamics (i.e.
radial structure, velocity shear etc.); the conversion of magnetic energy
into heat (i.e. magnetic reconnection); and interaction with the ambient
medium, leading to back flow, vortices and jet pinches (see $\SWa$).

The observed synchrotron emission from jets results from a line-of-sight
integration of the synchrotron emissivity, which depends on the (probably
complicated) field geometry, field strength and the distribution of the
relativistic electrons. Therefore, based on the observations alone it is not
possible to unravel the underlying field configuration.  We make a
sophisticated guess on how the magnetic pressure is distributed between the
entangled field and the ordered (azimuthal, poloidal or helical) magnetic fields.
Our heuristic method is explained in detail in Appendix~\ref{sec:MagneticField}.

For the quantity $\mathcal{N}$, related to the density of relativistic particles,
we use and compare two different models. The first model, which we shall refer
to as the \emph{\NTracer} model, makes extensive use of the tracers that are
advected with each jet component. This model takes into account the energy losses
due to adiabatic expansion, but neglects the effect of synchrotron cooling.
The second model is referred to as the \emph{\NCooling} model and is based on the
work of \citet{CamusThesis}, Section 4.3.4, and \citet{DelZanna2006}. It takes
into account both the effects of adiabatic losses and synchrotron cooling. There,
a power law of relativistic particles is injected along with the jet,
characterised by a number density of relativistic particles, $n$, and a cut-off
frequency $\nu_{\infty}$. The advantage of this model is that brightness
variations are given more accurately at all observer frequencies, but this model
is not able to separate the various contributions of different jet components. In
the case where the observation frequency is much lower than the cut-off frequency
(\mbox{$\nu_{\rm obs}<<\nu_{\infty}$}), there is no significant synchrotron
cooling. In this limit these two models should yield similar results, as we will
show. The general expression for the emissivity \equref{eq:jnuGeneral} will be
further specified in the next section for the \NTracer and the \NCooling models,
of which a full derivation can be found in the appendices
\ref{sec:RelativisticParticles},  \ref{sec:MagneticPfromGasP} and 
\ref{sec:MagneticField}.

\section{Method}
\label{sec:Method} 

We have generated $2.5D$ simulations of cylindrically symmetric jets, based on
similar simulations that we used in $\SWa$ and $\SWb$. For a complete explanation
of the numerical method that was used for generating the simulations, we refer
the reader to those papers. 

We employ the code MPI-AMRVAC
(see for example
\citealt{Keppens2012};
\citealt{Porth2014a};
\citealt{Xia2018} or
\citealt{Teunissen2019} and \url{amrvac.org}),
and use the special relativistic hydrodynamical
module (SRHDEOS) with the Mathews approximation for the Synge gas equation of
state, suitable for a non-relativistic gas, as well as a relativistically hot
gas. Moreover, we use the Harten-Lax-van Leer-Contact solver (HLLC)
(\citealt*{Toro1994}; \citealt{Migone2005}) combined with a Three step Runge-Kutta
time-discretisation scheme and a Koren limiter \citep{Koren1993}.

The base-level computational domain contains (120 $\times$ 480) grid cells,
corresponding to the physical scale of (120 $\times$ 480 kpc$^2$). We allow for
three additional refinement levels, resulting in an effective resolution of
(960 $\times$ 3840) grid cells. The jet is injected along the $Z-$axis at $Z=0$,
between $R = 0$ and $R = R_{\rm jet}(Z=0) = 1$~kpc.
We study jets with three different radial profiles, namely a piecewise isochoric
jet (model $A2$); an isothermal jet (model $I2$); and a homogeneous
jet (model $H2$). In this context, piecewise isochoric means that as the jet is
injected into the system, the jet spine and the jet sheath are initiated with
constant, but different mass densities. The jet spine is assumed to have a
lower mass density than the jet sheath. We assume that at the jet inlet the
jet spine and the jet sheath have already gone through some extent of turbulent
mixing, compared to the conditions close to the AGN, so that their mass ratio is
moderate. In correspondence with the previous papers $\SWa$ and $\SWb$, we have
arbitrarily chosen the mass density ratio to be $\rho_{\rm sh}/\rho_{\rm sp} = 5$.

The jets are switched on for 15.3 Myr; interrupted for 0.68 Myr; and
restarted again for another 6.8 Myr, matching realistic time scales for DDGRs.
For the jet in the initial phase (or simply ``the first jet"), and the jet in the
restarted phase (simply the ``second jet") different tracers are advected. For the
homogeneous jet ($H$) this leads to two different tracers, but in the case of
the radially structured jets (the $A$ or $I$ models) where we have a jet
spine and a jet sheath, a total of four tracers is used in each simulation.
In addition to the SRHDEOS module that was used for the simulations in $\SWa$ and
$\SWb$, a new set of transport equations was added to the code in order to trace
the number density and cut-off energy of a population of relativistic particles
with a power law distribution of the form $\epsilon^{-s}$, see
\equref{eq:TransportEquation_n};
\equref{eq:TransportEquation_n0} and
\equref{eq:TransportEquation_epsinf}.

The data files that are generated in the numerical simulations are used
in the relativistic ray-tracer synchro.py (e.g. \citealt*{Porth2014b})
in order to create the synthesized synchrotron maps.
By reflecting the one-sided jet in the $Z\: =\: 0$
plane (of the initial computational box), this jet is transformed into a
two-sided jet. The emissivity is then calculated in each grid cell of the new
box. We choose the size of the new box to be $700\times 700\times 700$ kpc$^3$
with a resolution of $300\times 300\times 300$ grid cells.

For $\vartheta=0^{\circ}$ the line of sight and the jet $Z-$axis are aligned
(head-on), while for $\vartheta=\pm 90^{\circ}$ the line of sight is
perpendicular to the jet axis (face-on). In our coordinate system positive
viewing angles correspond to the upper jets (the NW jets in Fig.\ref{fig:1}A)
pointing towards the observer, while the lower jets (the SE jets in
Fig.\ref{fig:1}A) point away from the observer. For negative viewing angles
this is the other way around, and corresponds to case of J1835+6204.
We simulated the jet models with a large variety of viewing angles and find a
best match with J1835+6204 for $\vartheta=-71^{\circ}$, which we will use in most
cases, unless mentioned otherwise. The sky plane rotation angle is close to
$\delta=30^{\circ}$, but we will use $\delta=90^{\circ}$ is most cases in order
to favour the visibility of the synthesized synchrotron maps.

\halfskip
Here we compare the two synchrotron radiation models, the emissivity of the
\emph{N-tracer} model:

\be
j_{\nu} \: \propto \: \Sigma_{\rm A} \: \mathcal{D}^{2+\alpha} \:
\mathcal{N_{\rm A0}} \:
(B_{\perp}')^{1+\alpha} \:
(\nu)^{-\alpha} \:
\left(\frac{\theta'_{\rm A}\:\rho'}{\rho'_{\rm A0}}
\right)^{1+\frac{2}{3}\alpha} \; ,
\ee
where the sum runs over the various jet components (see appendix
\ref{subsec:TheNTracermodel}), and the emissivity of the \NCooling model:

\be
j_{\nu} \propto \mathcal{D}^{2+\alpha} \:
\mathcal{N}_0 \:
(B_{\perp}')^{1+\alpha} \:
(\nu)^{-\alpha} \:
\left(\frac{n'_{\rm e}}{n'_{\rm e\:0}}\right)
^{1+\frac{2}{3}\alpha} \:
\left(1 - \sqrt{\frac{\nu'}{\nu'_{\infty}}} \right)^{2\alpha-1} \; ,
\ee
(see appendix \ref{subsec:TheNCoolingmodel}). Here $B_{\perp}'$ is given defined
through \equref{eq:BmixPerp2}. After the intensity is calculated a Gaussian
smoothening is applied with a FWHM of 8.0 kpc to match a typical beam size of
a radio telescope such as the VLA in B or D configuration at 4.8 GHz,
with beam sizes of $\sim$ 1.14'' -- 1.71'' as in the case of J1835+6204
(see \citealt{Konar2012}).

The brightness levels of the synthesized synchrotron maps are constructed as
follows: After integrating the emissivity along the line of sight, the peak level
radio flux $I_{\nu, {\rm max}}$ is determined per image. We let this flux level
correspond to the peak flux level of the observed source J1835+6204.
We choose a dynamic range of 256 from the peak intensity and show contours spaced
by factors of two. This allows us to make a realistic com\-parison between the
various synthesized synchrotron maps.

\section{Results and Discussion}
\label{sec:ResultsAndDiscussion}
In this section we present the results of the two synchrotron radiation
models that we have used for our synthesis maps. As a reference source, we will
use the DDRG J1835+6204 observed at 4860.1 MHz (see Fig. \ref{fig:1}, panel A).
We will summarise the main features of J1835+6204 here.

J1835+6204 has two clearly distinct pairs of radio lobes: the outer
North-Western and the South-Eastern lobes, referred to as NW1 and SE1,
respectively; and  the inner North-Western and the South-Eastern radio lobes
(closest to the central engine situated at the plus-sign in the center of the
image), referred to as NW2 and SE2, respectively. The viewing angle of the source
is such that the SE lobes are pointing towards the observer and the NW lobes are
pointing away from the observer. The outer radio lobes are thought to be caused
by the first jet eruption; the space between the outer lobes and the inner lobes
correspond to a quiescent phase in the activity of the central engine, where a jet
was not injected; and the inner radio lobes are due to a second jet eruption.
All four hotspots are visible and are well described by a power law, meaning
that the hotspots are still being fed by the jets (or the time passed since the
last material from the jet has reached the outer hotspot is relatively short).
Based on the distance of the hotspots to the central engine and the jet-head
propagation speed, it is possible to estimate the viewing angle of the radio jet
(see for example \citealt{Safouris2008} or \citealt{Konar2013b}).
We will briefly sketch this method here:
When a radio galaxy ejects radio jets of which the jet axis is not exactly
perpendicular to the line of sight, retardation effects come into play. The time
it takes for light emitted by (for example the hotspot of the) approaching jet
to reach the observer will be less than the time it takes for light emitted by
the (hotspot of) receding jet, when we assume the jets to have equal physical
size. This leads to a difference in the observed size of both jets.
When the projected size of the approaching jet is $d_{\rm a}$, and
the projected size of the receding jet is $d_{\rm r}$, the arm length ratio $D$
is then defined by:

\be
D \:=\: \frac{d_{\rm a}}{d_{\rm r}} \: = \:
        \frac{1 + \beta_{\rm hd}\cos{\vartheta}}
             {1 - \beta_{\rm hd}\cos{\vartheta}}\; ,
\ee
with $\beta_{\rm hd}$ the jet-head advance speed, and $\vartheta$ the viewing
angle, as before. This can be rewritten as:
\be
\cos{\vartheta} = \frac{1}{\beta_{\rm hd}}\frac{D-1}{D+1} \; .
\ee
The projected linear sizes $d_{\rm a}$ and $d_{\rm r}$ for both the outer jets,
as well as the inner jets for J1835-6204 are known (see \citealt{Konar2012}),
and lead to an arm length ratio of $D \approx 1.02$.
Moreover, the jet-head advance speed of the outer jets is typically of the
order $\beta_{\rm hd} \sim 0.03$ -- $0.05$. The jet-head advance speed for the
inner jets is usually much higher, and of the order
$\beta_{\rm hd} \sim 0.1$ -- $0.5$.
Both cases agree very well with our
simulations, as is discussed in $\SWb$. Using these values, we find an estimate
for the viewing angle $70^{\circ}\lesssim\vartheta\lesssim 89^{\circ}$.
The inner hotspots are brighter by a factor of 2 compared to the outer ones for the
NW jets and by a factor of 4 for the SE jets. The outer radio lobes are
more elongated, while the inner radio lobes are more circular.

  \subsection{Comparing J1835+6204 to models A2, I2 and H2}
  \label{subsec:SourceAndA2I2H2}

In $\SWa$ three radial jet profiles were explained in detail, namely:
\begin{itemize}
\item the homogeneous radial jet profile, denoted as `$H$'. This model describes
a non-rotating structureless jet profile;
\item the (piecewise) isothermal radial jet profile, denoted as `$I$'. This model
describes a jet that is initiated with an azimuthal velocity component, and
constant temperature $T_{\rm sp}$ throughout the cross-section of
the jet spine and constant (but possibly different) temperature $T_{\rm sh}$
throughout the jet sheath. We have chosen $T_{\rm sp}$ and $T_{\rm sh}$ to be
equal in these simulations;
\item and a (piecewise) isochoric jet profile denoted, as `$A$'. This model
describes a jet that is initiated with an azimuthal velocity component, and
constant (but different) mass densities throughout the cross-section of the
jet spine and the jet sheath.
\end{itemize}
In $\SWb$ the same radial profiles were used, but instead of the jets being
continuously injected, there is a small period of intermittency where the central
engine is not active. This interrupted jet flow leads to the behaviour as is
observed in DDRGs. We refer to these jet models as $H2$, $I2$ and $A2$,
respectively.

    \subsubsection{Similarities between the models A2, I2 and H2}
    \label{subsubsec:SimilaritiesA2I2H2}

\begin{figure*}
$
\begin{array}{@{\hspace{0pt}}c @{\hspace{20pt}}c}
\includegraphics[clip=true,trim=0.cm 0cm 5.2cm 7.6cm,width=0.41\textwidth]
{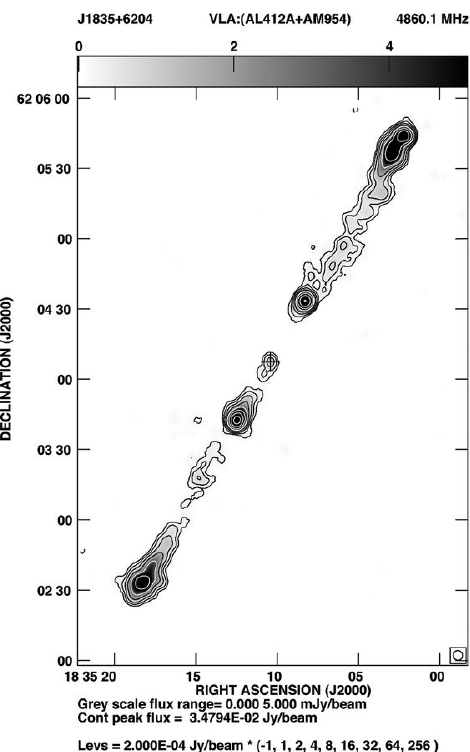}%
\put (-130,65) {\Large SE1}%
\put (-90,135) {\Large SE2}%
\put (-120,190) {\Large NW2}%
\put (-80,260) {\Large NW1}%
\put (-210,10) {\huge$\displaystyle A$}&
\includegraphics[clip=true,trim=0cm 0cm 0cm 0cm,width=0.45\textwidth]
{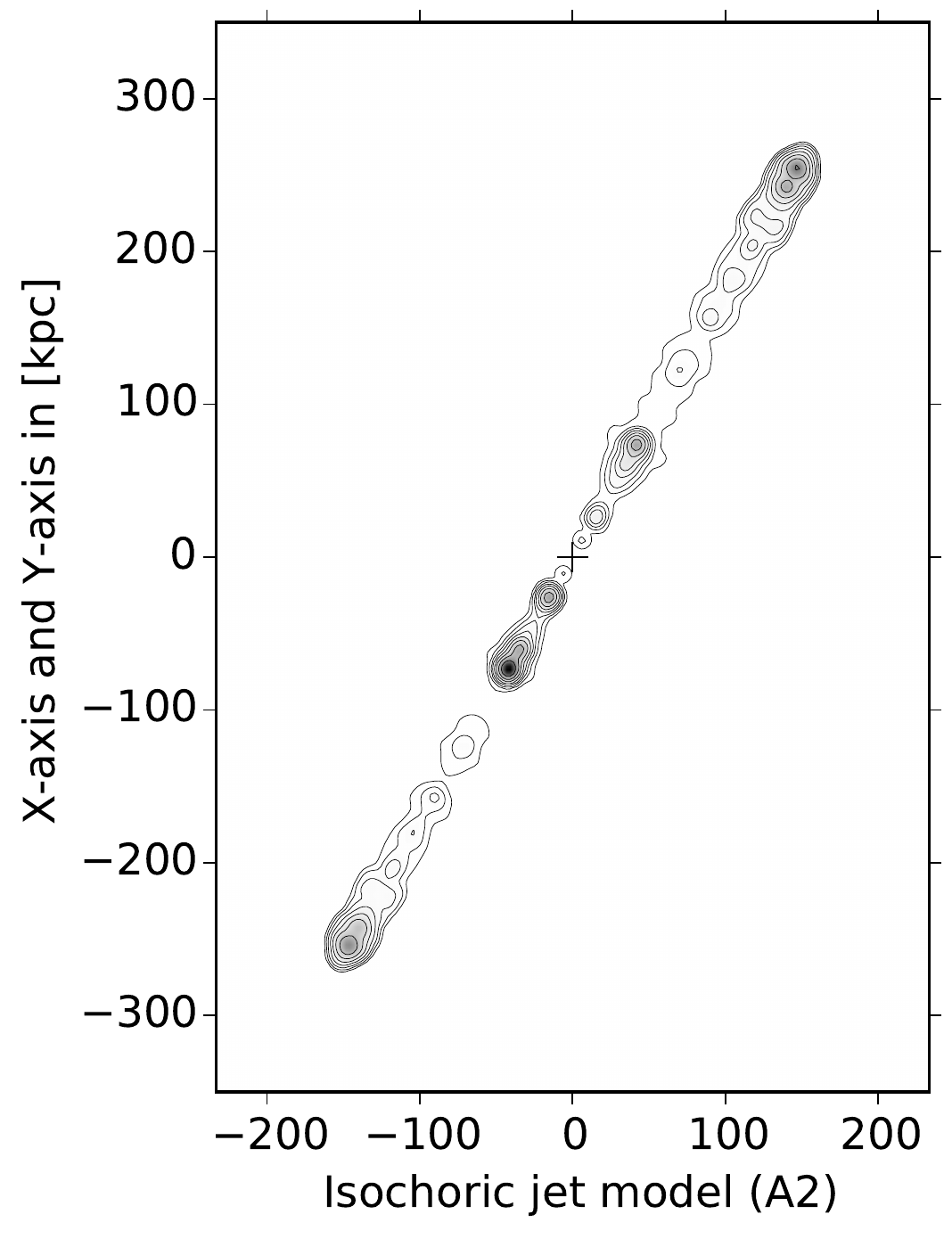}
\put (-210,10) {\huge$\displaystyle B$}

\end{array}
$

\caption{Comparison between the observed DDRG J1835+6204 and our best fit
synthesized jet model $A2$.
Panel A: The source J1835+6204 observed at 4860.1 MHz.
The cross-hairs correspond to the AGN radio core at the center of the radio
galaxy. (Credits: \citealt{Konar2013b}, Fig. 2, p5).
Panel B: The (piecewise) isochoric spine--sheath jet model ($A2$). The sky
plane rotation angle \mbox{$\delta = 30^{\circ}$}; the viewing angle
\mbox{$\vartheta = -71^{\circ}$}; a faction of 70 per cent of the magnetic
pressure contributed by entangled field \mbox{$\Lambda = 0.7$}; using the
\NCooling model at observation frequency
\mbox{$\nu_{\rm obs} = 6 \cdot 10^{-4} \nu_{\infty}$}.
The contour levels for the
synthesized synchrotron map have been chosen similar to the contours for the
observed source J1835+6204. The flux levels are
\mbox{$I_{\nu} \in \{1,2,4,8,16,32,64,256\} \times I_{\nu, {\rm min}}$}, with
\mbox{$I_{\nu, {\rm min}} = I_{\nu, {\rm max}}/256$} and
\mbox{$I_{\nu, {\rm max}} = 1$}.
The cross hairs correspond to the jet-inlet of the simulation, located at
a distance of 57 kpc from the AGN for a jet with a $1^{\circ}$
opening angle, but at the inlet we choose the opening angle to be $0^{\circ}$.
}
  \label{fig:1}
\end{figure*}

In Fig. \ref{fig:1} A, the radio image of J1835+6204 is compared
to our best fit synthesized radio maps of jet models Fig. \ref{fig:1} B:
The (piecewise) isochoric jet model $A2$; with
sky plane rotation angle \mbox{$\delta = 30^{\circ}$}; using the
\NCooling synchrotron radiation model at an observation
frequency \mbox{$\nu_{\rm obs} = 6 \cdot 10^{-4} \: \nu_{\infty}$}; with
70 per cent of the magnetic pressure contributed by the entangled fields
(\mbox{$\Lambda = 0.7$}) and 30 per cent contributed by ordered helical
fields (pitch angle \mbox{$\kappa = 45^{\circ}$}); and
viewing angle \mbox{$\vartheta = -71^{\circ}$}.

In Fig. \ref{fig:2} A, B and C, jet models $A2$; $I2$ and $H2$ are shown
respectively, with $\delta = 90^{\circ}$ to improve visibility; a purely
entangled magnetic field configuration ($\Lambda = 1$), and using the
\NTracer synchrotron radiation model at \mbox{$\vartheta = -71^{\circ}$}.
The interruption time for all three models is chosen to be 4.5 per cent of the
injection time of the initial jet, in correspondence with J1835+6204
which is estimated to have an interruption of $\sim$ 1~Myr and an active phase of
the previous outburst of $\sim$~22~Myr.
The observations of J1835+6204 show that the ratio of the distance from the
outer jet-heads to the AGN ($D_{\rm jt_1}$) to the distance from the inner
jet-heads to the AGN ($D_{\rm jt_2}$) is approximately
$D_{\rm jt_1} / D_{\rm jt_2} = 7\: :\: 2$ (at $\nu_{\rm obs} \sim$ 4886 MHz
and a beam size of $\sim$ 1.4''). Our simulations match this ratio best at a
time of 16.6~Myr, when the restarted jets have been injected for $\sim 0.6$~Myr.

All three synthesized synchrotron maps globally show the same behaviour:
the two pairs of hotspots are clearly visible at the termination shocks of
the jets and counter jets, and a few additional radio blobs can be found along
the jet axis. In all three models, the ratio between the distance from the central
engine to the inner hotspots ($D_{\rm hd2}$) and to the outer hotspots
($D_{\rm hd1}$) is approximately the same:
$D_{\rm hd1}/D_{\rm hd2} \approx 2 - 3$.
The maximum intensity of the SE2 hotspots is clearly higher than that of the NW2
hotspots, whereas the NW1 hotspots in all three models are brighter than the SE1
hotspots. Finally: the bulk of the jet flow is not visible. Most of the emission
comes from the hotspots, the cocoon surrounding the jet and some knots along the
jet axis associated with instabilities in the jet/cocoon interface. At this dynamic
range in intensity, the radio structure is broken up into distinct patches. Of the
three jet models, the morphology of the isochoric jet model A2 compares best to the
observed source J1835+6204. 
%
%
    \subsubsection{Differences between the models A2, I2 and H2}
    \label{subsubsec:DifferencesA2I2H2}

The main differences between the three models are:
[1] the shape of the (outer) hotspots;
[2] the width of the surrounding cocoon (the radio lobes) and
[3] the intensity contrast between the inner two hotspots and the outer two hot
spots. The isochoric jet model $A2$ shows almost round jet hotspots in both
the inner and outer jets.

The outer hotspots of the isothermal jet model $I2$ and of the homogenous jet
model  $H2$ appear to be more flattened in the direction perpendicular to the
jet flow. The inner- and outer jets in the $I2$ and $H2$ models have a distinct
morphology and are clearly separated in the synthesized synchrotron map.
The inner jets of the $A2$ jet model, on the other hand, lie partially within
the radio structure of the outer jet radio structure, in good agreement with
J1835+6204. In all cases we find this typical distinction between the $A2$,
the $I2$ and the $H2$ jet models.

  \subsection{Influence of the radial profile of the jets}
  \label{subsec:InfluenceOfRadialProfile} 

Our previous studies ($\SWa$ and $\SWb$) focussed on the dynamical differences
between the (piecewise) isochoric jet ($A$) model, the isochoric jet
$I$ model and the structureless homogeneous jet $H$ model. A strong structural
integrity means that the jet is not easily affected by pressure fluctuations
from the back-flowing jet material in the cocoon that surrounds the jet due to
the formation of vortices, so that internal shocks along the jet axis
are much less capable of disrupting the jet flow. Since the emissivity depends on
the magnetic field strength, which in turn depends on the gas pressure for an
entangled magnetic field configuration, radio features along the jet axis will
become more apparent for jets with less structural integrity. This agrees well
with the results in Fig. \ref{fig:1} B and Fig. \ref{fig:2} A.. 

Our simulations show that the $H2$ jet and the $I2$ jet maintain their structural
integrity well: interaction with the cocoon does not lead to a significant
disruption of jet flow. In contrast: the $A2$ jet suffers most disruption. This
has a large influence on the radio morphology of these jets, as we discuss in
the following subsections.

    \subsubsection{The homogeneous jet model, the outer jets}
    \label{subsubsec:HomogeneousJet} 

The homogeneous jet has the strongest structural integrity of the three models.
Therefore, the homogeneous jet is less easily deformed in radial direction, so
the jet flow will remain relatively close to the jet axis, in contrast to what
happens in jets with a well-defined spine/sheath structure. This causes the Mach
disc (termination shock at the jet-head) of the homogeneous jet to be less
diffuse and more flat than the Mach discs of the structured jets, which have a
larger surface area (the effective impact area) and have a bowl-shaped,
elongated structure  (see $\SWa$ for more details on these jet-head structures).
Therefore, the shock-heated gas for the homogeneous jet will be more concentrated
at the jet-head and the synchrotron radiation generated in this part of the jet
will outshine the other parts of the jet/cocoon structure.

\begin{figure*}
$
\begin{array}{c}
\includegraphics[clip=true,trim=0cm 0cm 0cm 0cm,width=0.8\textwidth]
{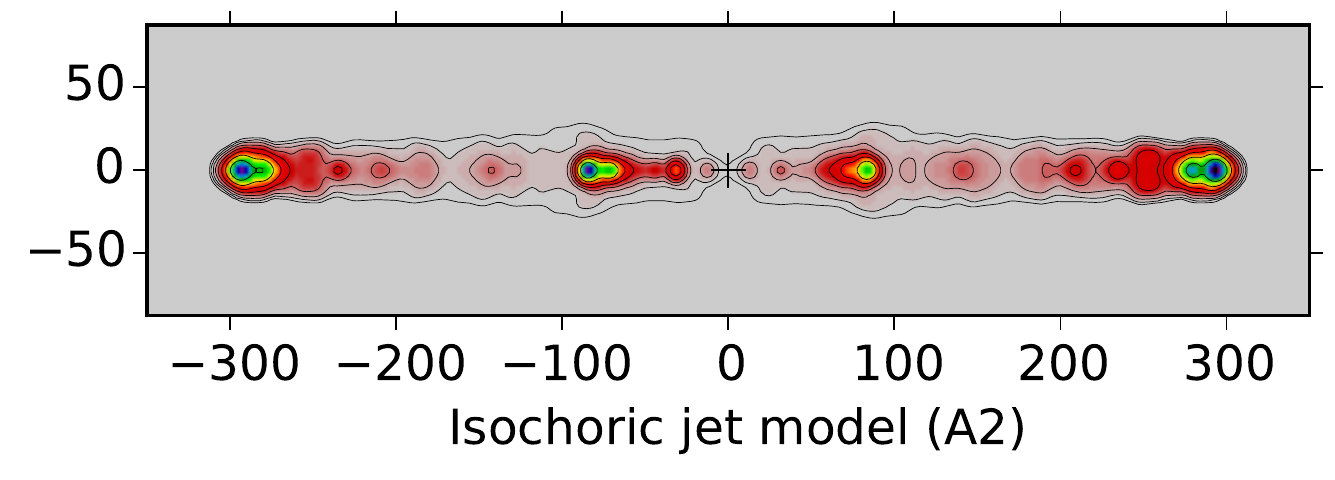}
\put (-420,30) {\huge$\displaystyle A$}\\
\includegraphics[clip=true,trim=0cm 0cm 0cm 0cm,width=0.8\textwidth]
{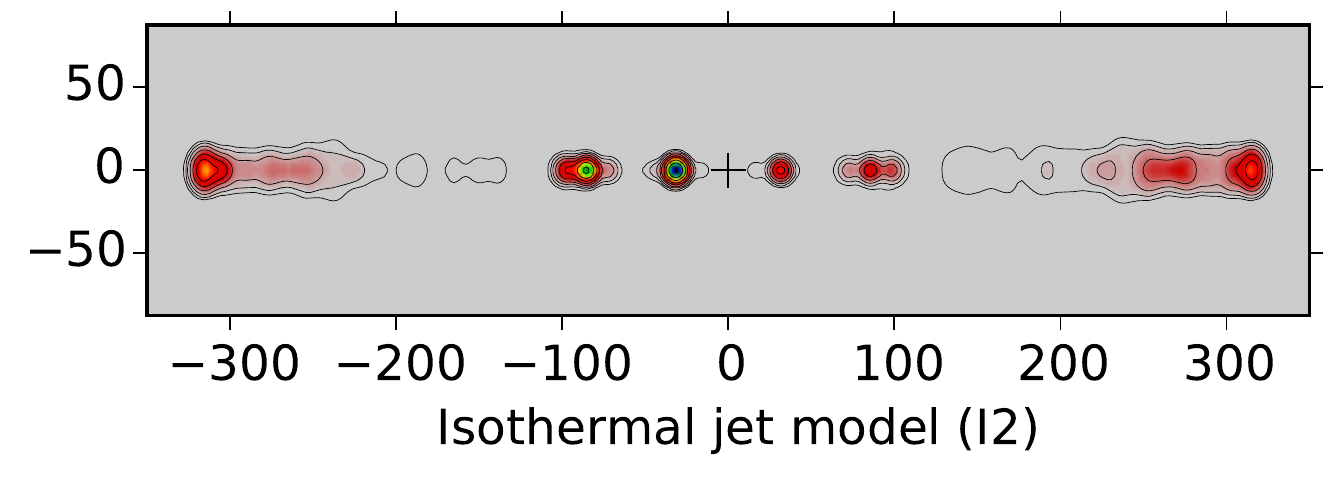}
\put (-420,30) {\huge$\displaystyle B$}\\
\includegraphics[clip=true,trim=0cm 0cm 0cm 0cm,width=0.8\textwidth]
{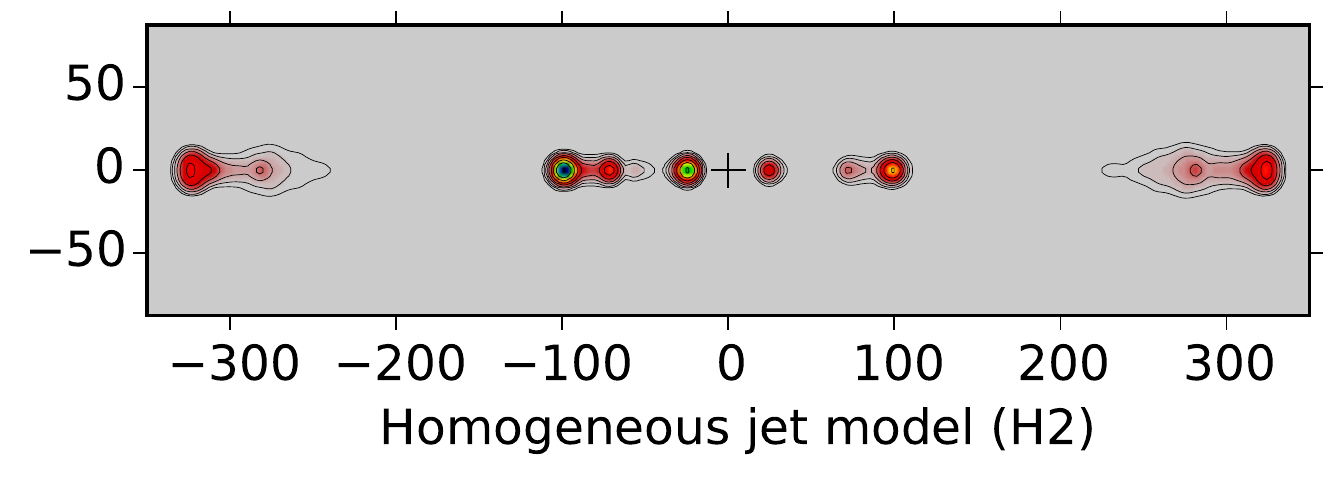}
\put (-420,30) {\huge$\displaystyle C$}
\end{array}
$
%
\caption{
Panels A, B and C show the jet models $A2$, $I2$ and $H2$ with the sky
plane rotation angle \mbox{$\delta = 90^{\circ}$} and a continuous color palet
for better visibility compared to Fig. \ref{fig:1} B. For these images, the
\NTracer model was used, with viewing angle \mbox{$\vartheta = -71^{\circ}$}
and a purely entangled field configuration, corresponding to \mbox{$\Lambda = 1$}.
Panel A: The synthesized synchrotron map of the isothermal spine--sheath jet
model ($A2$).
Panel B: The synthesized synchrotron map of the isothermal spine--sheath jet
model ($I2$).
Panel C: The synthesized synchrotron map of the homogeneous structureless jet
model ($H2$).
The flux levels are
\mbox{$I_{\nu} \in \{1,2,4,8,16,32,64,256\} \times I_{\nu, {\rm min}}$}, with
\mbox{$I_{\nu, {\rm min}} = I_{\nu, {\rm max}}/256$} and
\mbox{$I_{\nu, {\rm max}} = 1$}.
}
  \label{fig:2}
\end{figure*}
%

As the jet material propagates from the jet inlet to the jet-head, it
passed through $\sim$ 9 internal shocks at this length scale for models $A2$,
$I2$ and $H2$. These internal shocks are caused by jet-cocoon interactions:
pressure fluctuations in the cocoon for example by vortices that are created at
the jet-head structure and flow away from the jet head, as was also discussed in
$\SWa$. From the simulations in $\SWa$, a trending increase in temperature,
pressure and relativistic heat (through the effective polytropic index) was
found along the jet axis, clearly correlating to the passing of the jet material
through internal shocks. From these simulations we found that the jet material
was shocked to near-relativistic temperatures just ahead of the termination shock
at the jet-head, and shocked to relativistic temperatures just after passing the
termination shock. This explains the synchrotron intensity contours of the outer
jets in Fig. \ref{fig:2} C: bright outer hotspots and no radio features from the
outer cocoon, and small knots just ahead of the jet-heads.

    \subsubsection{The spine--sheath jet models $I2$ and $A2$: the first outer jets}
    \label{subsubsec:SpineSheathJets} 

The (piecewise) isothermal jet model $I$ assumes a constant tem\-pera\-ture across
a radial cut of the jet, but allows for the temperature of the jet spine and jet
sheath to differ mutually. However, in order to minimise the number of free
parameters, we've chosen the jet spine and sheath to have an equal temperature at
the jet inlet, \mbox{$T \sim 2 \times 10^9$ K}, which results from our choice of
parameters (such as number density, kinetic luminosity of the jet, Lorentz factor
of the bulk jet material), typical for FR-II jets and their ambient medium (see
$\SWa$ for a more detailed discussion). After passing the first internal shock,
the temperature of the jet material is increased to
\mbox{$T \sim 10^{10}$ -- $10^{11}$ K}, in agreement with observations (e.g.
\citealt{Homan2006} or \citealt{Kellerman2007}). This choice results in a
continuous mass density profile at the jet spine/sheath interface.
The isochoric jet model $A$, on the other hand, has a mass density contrast of
a factor 5 at the spine--sheath interface at the jet inlet. Therefore, the
isothermal jet is much more stable against the effect of pressure variations
within the cocoon (from back flowing vortices) than the isochoric jet.
As a result, the isothermal jet maintains more of its structural integrity than
that of the (piecewise) isochoric jet as one moves further away from the central
engine. The jet-head of the isochoric jet has a very elongated and bowl-shaped
structure. Therefore, the region where the jet flow is shock-heated is larger
and more diffuse, causing the radio structures of the jet-head (and surrounding
back-flowing material) to be more extended than in the homogeneous and isothermal
jet models. This explains the synchrotron intensity contours of the outer jets in
panels B and C of Fig. \ref{fig:2}: a relatively flat jet
head from the isothermal jet, with radio features from the back-flowing
material (and the Doppler boosting of back-flowing material is actually quite
prominent); while the jet-head from the (piecewise) isochoric jet is more round
and the surrounding back-flowing material is much better observed.

    \subsubsection{The restarted, inner jets}
    \label{subsubsec:InnerJets} 

The inner (restarted, or second) jets of all three jet models $A2$, $I2$ and
$H2$ propagate through a much more dilute external medium: the disturbed
intergalactic medium. It consists of a mixture of shocked jet material and
shocked intergalactic material, with temperature, pressure and mass density
comparable to that of the jet itself. The strength of the termination shock at
the jet-head is determined by the (relativistic) Mach number: the ratio of the
(relativistic) velocity of the jet-head as measured in the observer frame, and
the (relativistic)  sound velocity of the ambient medium (see $\SWb$ for more
details). This ratio will be significantly less for the restarted jets, because
of the much higher temperature and low mass density of the material left
behind by the first jet, so the Mach disc of the inner jets will not be as strong.
The Mach number for the Mach disc of the outer jets for jet models
$A2$, $I2$ and $H2$ for Fig.\ref{fig:2} are \mbox{$\sim$ 13, 22 and 13}
respectively, while the Mach numbers for the Mach disc of the corresponding inner
jets are \mbox{$\sim$ 8, 16 and 11}, indeed a factor of 0.6 -- 0.8 as small for
all three jet models. This is why the inner hotspots will be relatively less bright
and do not outshine the surrounding jet material to the same extent as the hotspots
in the first (older) jet that impact the intergalactic gas (this is why these inner
jet-head radio features are also referred to as `warm spots', see for example
\citealt{Konar2012}).
The general characteristics of these three jet models (as described in the
previous section) will remain to be true, but more of the jet material along the
jet axis will be visible, as compared to the first (older) jets.
Finally, the jet-head advance speed strongly depends on the mass density
ratio between jet material and ambient medium material,
\mbox{$\eta_{\rm R} = \rho_{\rm jt} / \rho_{\rm am}$}. A small $\eta_{\rm  R}$
(as is the case for jets that are under-dense compared to their ambient medium)
leads to a smaller jet-head advance speed, while $\eta_{\rm R} \sim 1$ (for jets
that have mass density comparable with that of their ambient medium,
\mbox{$\rho_{\rm jt} \approx \rho_{\rm am}$)} leads to a higher jet-head advance
speed (see $\SWb$). The inner jets in the models $A2$, $I2$ and $H2$ have mass
density ratio very close to \mbox{$\eta_{\rm R} = 1$}, so they propagate much
faster through the cocoon that was left behind by the first jets, than the
propagation speed of the first jets themselves, which have
\mbox{$\eta_{\rm R} \approx 0.01$}, despite their lower Mach numbers. Therefore,
the discharge of jet material through the Mach disc of the inner jets will be
much smaller, leading to a back-flow that is also less strong. As a result, the
inner jets encounter less pressure variations, and maintain their integrity
better than the outer jets. The structured spine--sheath jets have a higher
Lorentz factor jet spine. In case of the inner jets, this high Lorentz factor jet
material will be able to propagate all the way up to the jet-head. This is why
the effect of Doppler (de-)boosting is so clearly seen for the inner jets in all
three models. In this work we are trying to explain the radio morphology of a DDRG
such as J1835+6204. We find that the only possible jet model that reproduces
its resolution-frequency specific view from Fig.~\ref{fig:1}, is the (piecewise)
isochoric jet model $A2$. Therefore, we continue with this model for further
examination of the various parameters.
%
\begin{figure*}
$
\begin{array}{c}
\includegraphics[clip=true,trim=0cm 0cm 0cm 0cm,width=0.8\textwidth]
{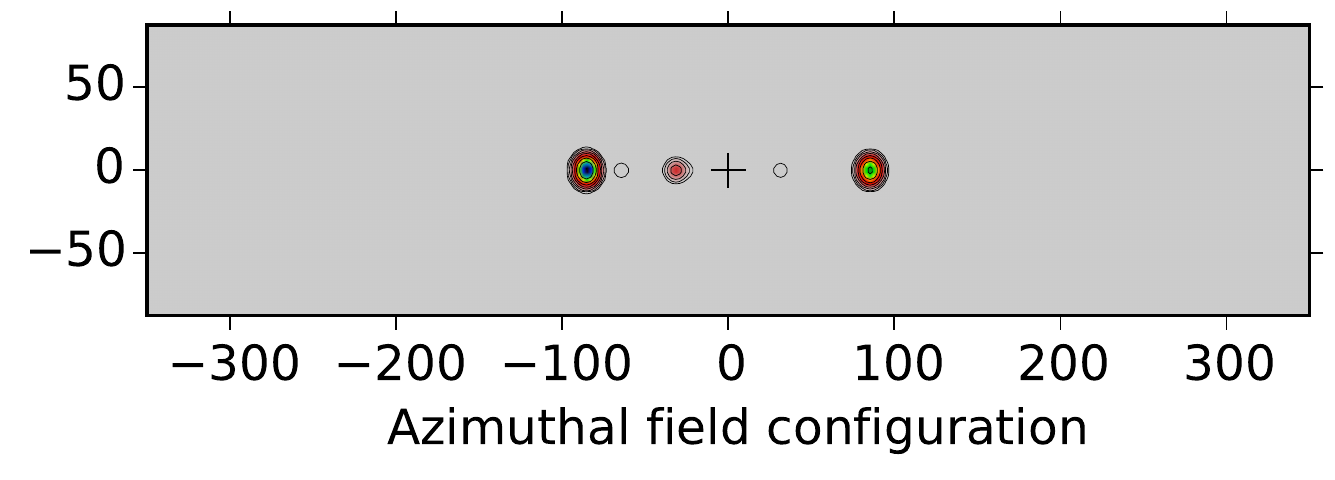}%
\put (-420,30) {\huge$\displaystyle A$}\\
\includegraphics[clip=true,trim=0cm 0cm 0cm 0cm,width=0.8\textwidth]
{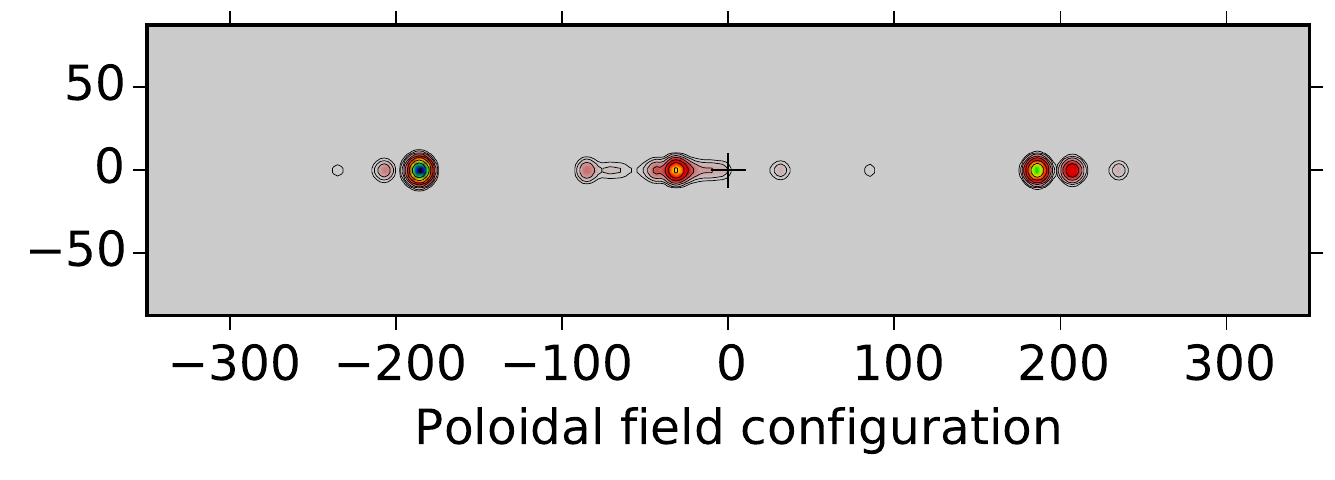}
\put (-420,30) {\huge$\displaystyle B$}
%
%
\end{array}
$

\caption{{\bf Purely} ordered magnetic field configurations for the structured
isochoric jet model $A2$.
The entangled magnetic fields are absent and the ordered magnetic field are
assumed to exist only within the jets.
Panel A assumes a pure azimuthal field configuration
(so $\Lambda = 0$ and $\kappa = 0^{\circ}$);
Panel B assumes a pure poloidal field configuration
(so $\Lambda = 0$ and $\kappa = 90^{\circ}$);
The flux levels are
\mbox{$I_{\nu} \in \{1,2,4,8,16,32,64,256\} \times I_{\nu, {\rm min}}$}, with
\mbox{$I_{\nu, {\rm min}} = I_{\nu, {\rm max}}/256$} and
\mbox{$I_{\nu, {\rm max}} = 1$}.
}
  \label{fig:3}
\end{figure*}
%
%
%
\begin{figure*}
$
\begin{array}{c}
\includegraphics[clip=true,trim=0cm 0cm 0cm 0cm,width=0.8\textwidth]
{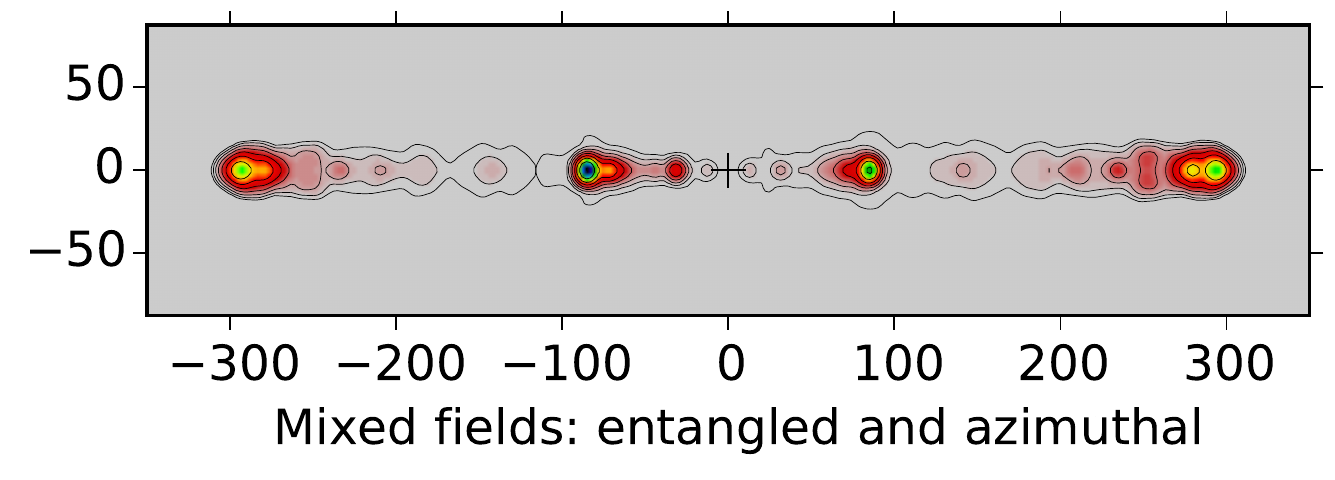}%
\put (-420,30) {\huge$\displaystyle A$}\\
\includegraphics[clip=true,trim=0cm 0cm 0cm 0cm,width=0.8\textwidth]
{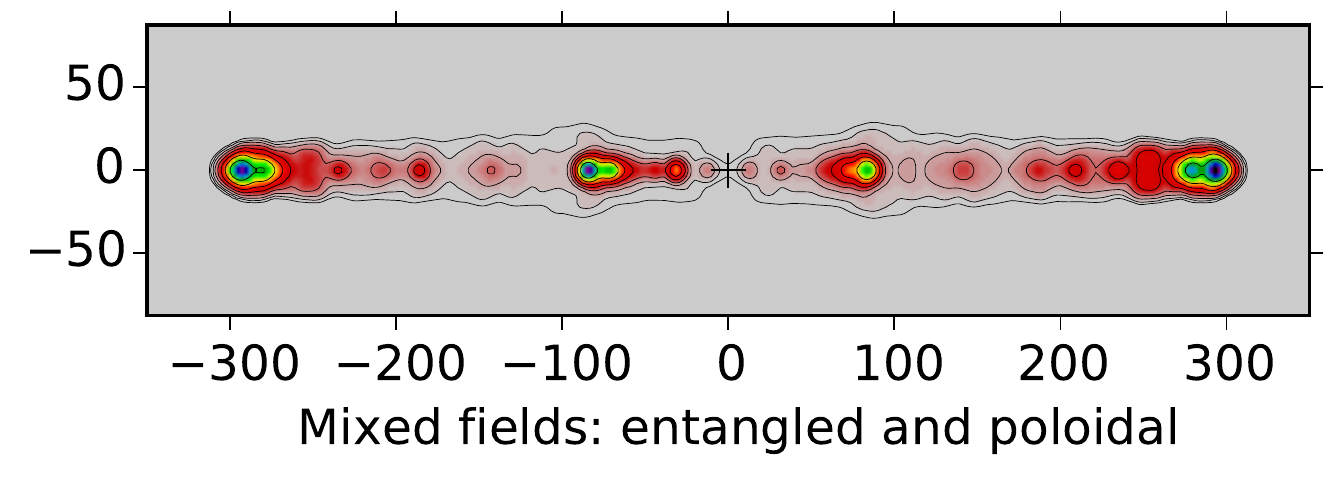}%
\put (-420,30) {\huge$\displaystyle B$}\\
\includegraphics[clip=true,trim=0cm 0cm 0cm 0cm,width=0.8\textwidth]
{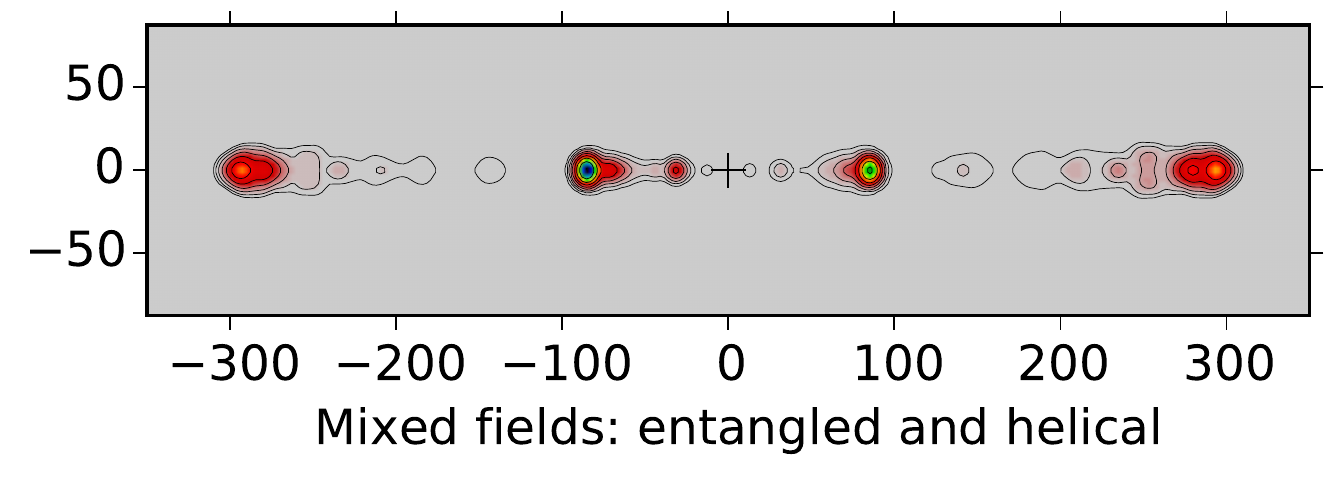}%
\put (-420,30) {\huge$\displaystyle C$}\\
\includegraphics[clip=true,trim=0cm 0cm 0cm 0cm,width=0.8\textwidth]
{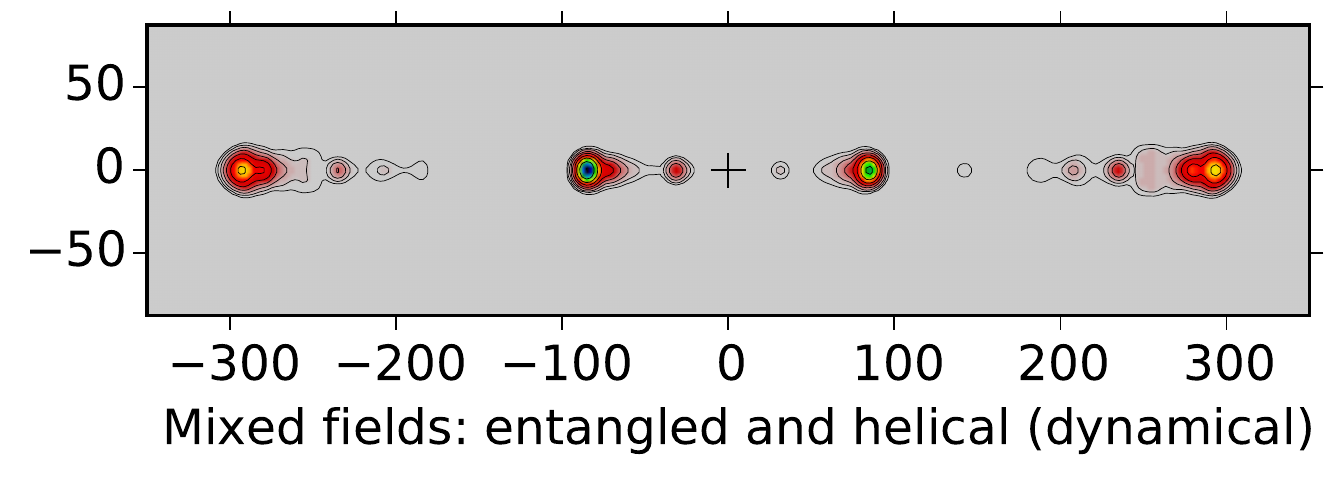}%
\put (-420,30) {\huge$\displaystyle D$}
\end{array}
$

\caption{Various mixed magnetic field configurations for the isochoric jet
model ($A2$): In panels A end B the magnetic pressure of the entangled fields in
the jets is half of the total magnetic pressure (so $\Lambda = 0.5$).
Panel A: within the jets, an azimuthal magnetic field configuration is assumed
(so $\kappa = 0^{\circ}$).
Panel B: within the jets, a poloidal magnetic field configuration is assumed.
(so $\kappa = 90^{\circ}$).
Panel C: within the jets, a helical magnetic field configuration is assumed
with $\kappa = 45^{\circ}$. The magnetic pressure of the entangled fields is now
chosen to be one quarter of the total magnetic pressure (so $\Lambda = 0.25$).
Panel D: also here a helical magnetic field configuration is assumed
with $\kappa = 45^{\circ}$, however, in this case we use the absolute mixing
factor $\Lambda_{\rm AB}$ for the fraction of the magnetic pressure.
The flux levels are
\mbox{$I_{\nu} \in \{1,2,4,8,16,32,64,256\} \times I_{\nu, {\rm min}}$}, with
\mbox{$I_{\nu, {\rm min}} = I_{\nu, {\rm max}}/256$} and
\mbox{$I_{\nu, {\rm max}} = 1$}.
}
  \label{fig:4}
\end{figure*}

%
  \subsection{Comparing various magnetic field configurations}
  \label{subsec:FieldConfigurations}

In this section we focus on the main differences between the different magnetic
field configurations and which magnetic field configuration(s) we expect to be most
realistic.
In the simulations at this dynamical range, the purely structured magnetic fields
appear as knots along the jet axis. Here, we have assumed the structured magnetic
fields outside the jets (within the cocoon) to be so small (in fact, absent)
compared to the fields inside the jet flow that they can be ignored.

The \emph{general} behaviour of the radio features caused by structured magnetic
fields within the jets can be explained as follows: For the sake of simplicity of
the argument we assume the Doppler factor to be small. Then the line of sight
vector measured in the observer frame, $\bm{n}$, and as measured in the plasma
rest-frame, $\bm{n'}$, are equal. In that case, the magnetic field component
perpendicular to the line of sight, $B_{\perp}$, is equal in both frames.
Then, when a radio jet is viewed face-on (so the line of sight is perpendicular
to the jet axis), the magnitude of the perpendicular field component of the
poloidal field is at its maximum, while the magnitude of the azimuthal field is
at its minimum. Therefore, one can expect a larger contribution to the
perpendicular field component coming from the poloidal field than from the
azimuthal field. The opposite is true when the jet is viewed head-on (so when the
line of sight vector coincides with the direction of the jet-axis). In most of the
synthesized synchrotron images in this paper, the jets have a viewing angle of
$\vartheta = -71^{\circ}$, tending toward a face-on situation. This might explain
why the synthesized synchrotron images from the poloidal field configuration are
more prominent than those of the azimuthal field configuration (as discussed in the
next alinea). In Fig. \ref{fig:3}, the structured jet model A2 is shown in the case
where the magnetic fields are purely ordered (\mbox{$\Lambda = 0$}, see appendix
\ref{subsec:Bmix}). Since we assume these ordered fields are confined to the jets,
the synchrotron emission has a very thin, elongated morphology. Panel A assumes a
purely azimuthal magnetic field (where the pitch angle \mbox{$\kappa = 0^{\circ}$}),
whereas panel B assumes a purely poloidal magnetic field (where the pitch angle
\mbox{$\kappa = 90^{\circ}$}). We note that for these images, the position of the
outermost knots in the synchrotron radiation is just ahead of the termination shock
of the jets (relative to the central engine). This means that in case of the purely
ordered fields, the termination shock itself does not contribute the most to the
radio emission. An explanation for this effect is that the ordered fields are
strongest where the jet radius is pinched the most. This occurs just ahead of the
termination shock due to high-pressure shock-heated back-flowing jet material in
the surrounding cocoon.

    \subsubsection{Structured magnetic fields for model A2}
    \label{subsubsec:StructuredMagneticFieldA2}

Fig. \ref{fig:3} A shows the A2 jet for a purely azimuthal
magnetic field configuration (\mbox{$\bm{B} = \bm{B}_{\phi}$}). Only the emission
from the restarted jets can be seen in this image, meaning that the contribution
from the initial jets is at least a factor of 256 smaller than the peak brightness
level in the image. The radio structure is not continuous along the jet axes, but
just shows a number of knots (three on each side). These knots are a result of
the jet being locally compressed by external pressure fluctuations. Such compressions
cause the jet radius $R_{\rm jt}$ to reduce and since
$B_{\phi} \propto R/R_{\rm jt}^2$, a local compression of jet radius causes a local
increase in magnetic field strength. As a result, the emissivity is also enhanced.
These local maxima in synchrotron emissivity appear as knots (for the dynamical range
in radio emission chosen in this paper). Each knot from the SE2 jet is brighter than
the corresponding knot on the NW2 jet. The Lorentz factor of these inner jets is
relatively high. Therefore, the brighting of the (approaching) SE2 radio features
compared to the corresponding (receding) NW2 jets is due to Doppler boosting/dimming.

Fig. \ref{fig:3} B shows the A2 jet for a purely poloidal
magnetic field configuration (\mbox{$\bm{B} = \bm{B}_{Z}$}). In contrast to the
azimuthal case, here the contribution from the initial jets is most clearly
visible. The contribution from the NW2 jet is virtually absent. The same argument
used to explain the appearance of knots in the azimuthal case holds for
the poloidal case, with the only difference that the magnitude of $B_{Z}$ scales as
$B_{Z} \propto 1/R_{\rm jt}^2$. The two knots that are
the brightest in the SE2 jet in the azimuthal case are also seen in the poloidal
case, but they are significantly weaker than outermost knots. The outer knots
from jet NW1 are slightly brighter than the corresponding SE1 jets.

    \subsubsection{Mixed magnetic fields for the model A2 with $\Lambda = 0.5$}
    \label{subsubsec:MixedMagneticFieldsA2Lambda0dot5}

Fig. \ref{fig:4} shows the images resulting from mixed turbulent (entangled) and
regular (ordered) magnetic field configurations (see \ref{subsec:Bmix} for more
information). Panels A and B show the images where the (maximum) energy of the
entangled field equals the (maximum) energy of the ordered magnetic fields,
corresponding to \mbox{$\Lambda = 0.5$}.

Panel A shows the case where the ordered fields are azimuthal, and panel B
shows the case where the ordered fields are poloidal.
The effect of adding the azimuthal field to the entangled field is that the inner
radio structure (from the restarted jets) and the outer radio structures (from the
initial jets) break up. The overall characteristics remain, however, the relative
brightness of the outer radio features is significantly less than in the case with
the purely entangled fields.
Adding a poloidal field to the entangled field leads to less obvious changes in
the synthesized map. The inner and outer radio structures do not break
up in this case and the brightness contrasts between inner and outer jets (NW2
v.s. NW1 and SE2 v.s. SE1) do not change significantly. Close examination shows that
knots between the hotspots become slightly brighter.

    \subsubsection{Mixed magnetic fields for the model A2 with helical fields}
    \label{subsubsec:MixedMagneticFieldsA2Lambda0dot25}

Panels C and D of Fig. \ref{fig:4} show the case of a mixed field
configuration, where the ordered magnetic fields are helical with a
fixed 
pitch angle \mbox{$\kappa = 45^{\circ}$}.
In panel C, the (maximum) magnetic energy from the entangled fields
is one third of the (maximum) magnetic energy from the ordered magnetic field, so
\mbox{$\Lambda = 0.25$}. In this case, the contribution from the initial jets
NW1 and SE1 to the total brightness becomes less significant and their radio
features become less elongated. Radio features between the inner hot
spots NW2 and SE2, and the outer hotspots NW1 and SE1 have almost vanished.
In panel D, a model is shown where the magnetic energy fraction
$\Lambda$ is not a fixed number between 0 and 1, but is actually equal to the
mass-weighted mixing factor $\Lambda_{\rm AB}$ (see $\SWa$ and $\SWb$). This
quantity calculates to what extent the spine and sheath of a structured jet
have mixed and therefore could be a measure for the amount of entanglement of the
magnetic fields.
Comparing this image to panel C, we notice quite
a lot of similarities with respect to brightness contrasts, shape of the radio
features, the number of knots, etc. Therefore, if the mass-weighted mixing factor
$\Lambda_{\rm AB}$ were to be used as a measure of the entanglement, it would
roughly correspond to \mbox{$\Lambda \sim 0.25$}.
If these simulations represent a realistic scenario, then the overall radio
morphology of a typical DDRG is clearly dominated by an entangled field
configuration. At the physical length scales that we are simulating, it is
reasonable to assume that the structured magnetic fields have been entangled
significantly due to turbulence and passing multiple internal shocks. The knots
as a result of a pinched jet radius for the structured magnetic fields appear just
before (so upstream of) the actual internal shocks along the jet axis, since the
jet radius has a local minimum there. The knots as a result of reduced jet radius
for the entangled magnetic fields on the other hand appear just after (so
downstream) of the internal shocks, since the gas pressure (which is assumed
to be close to equipartition with the magnetic pressure) has a local maximum.
Through the observation of such sources at radio frequencies, information about the
local field configuration can be gained, and can be checked with polarization
measurements.

  \subsection{Effects of spectral ageing}
  \label{subsec:SpectralAging}

Spectral ageing is a crucial aspect to consider, since radiative losses determine
whether or not one can see the outer lobes in radio maps, vital to know in the
area of wide area radio surveys (like LoTSS, \citealt{Shimwell2019}). Generally
speaking, spectral ageing of the relativistic electrons allows one to estimate the
age of the source, or - if no ageing is seen - to infer the need for re-acceleration
inside the jets or radio lobes.

\begin{figure*}
$
\begin{array}{c}
\includegraphics[clip=true,trim=0cm 0cm 0cm 0cm,width=0.8\textwidth]
{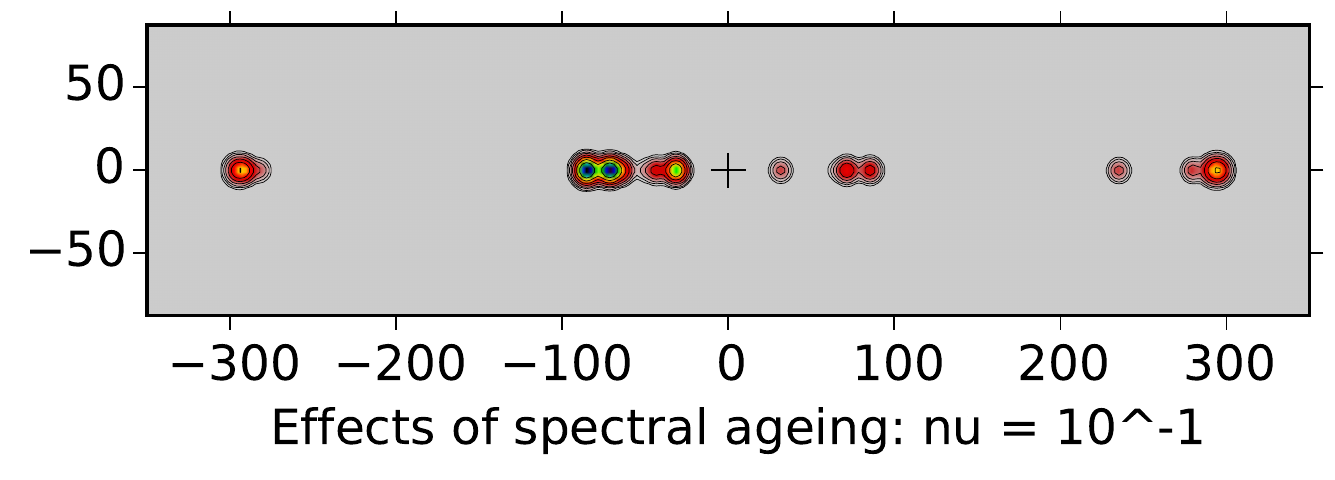}%
\put (-420,30) {\huge$\displaystyle A$}\\
\includegraphics[clip=true,trim=0cm 0cm 0cm 0cm,width=0.8\textwidth]
{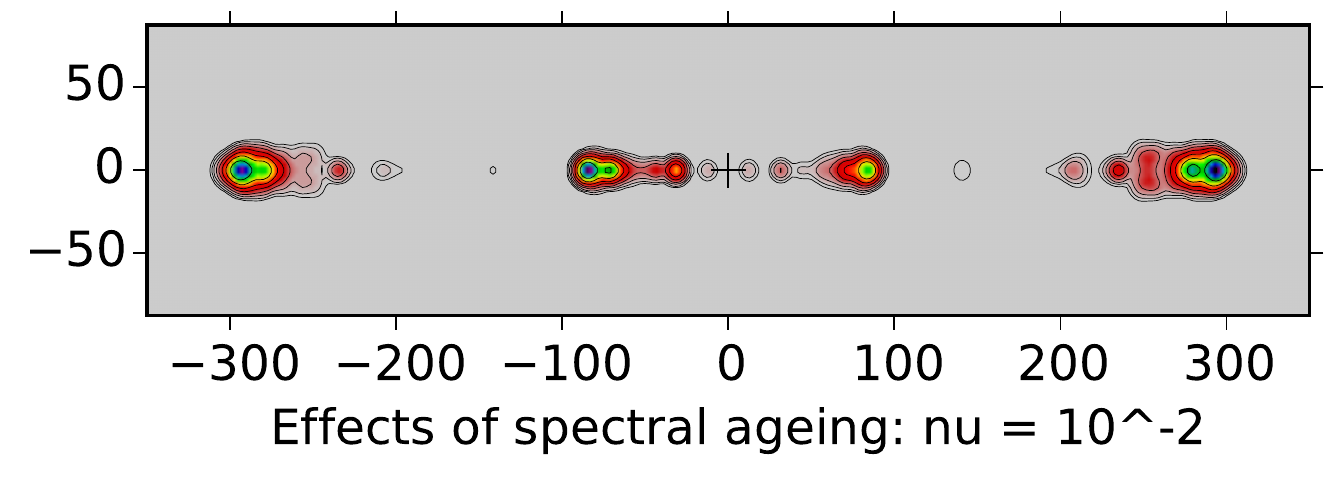}%
\put (-420,30) {\huge$\displaystyle B$}\\
\includegraphics[clip=true,trim=0cm 0cm 0cm 0cm,width=0.8\textwidth]
{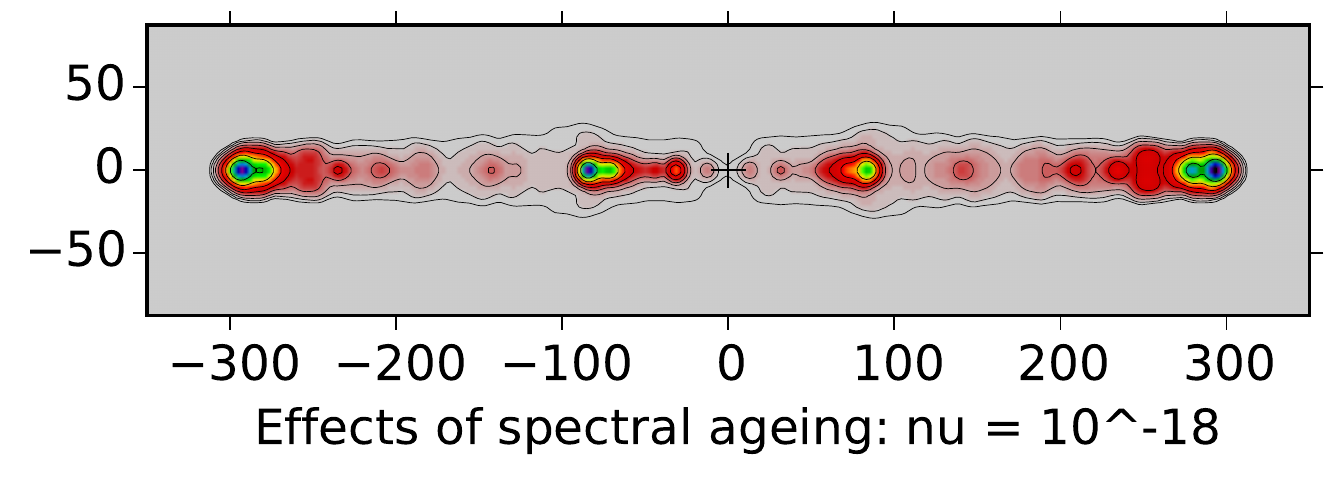}%
\put (-420,30) {\huge$\displaystyle C$}\\
\includegraphics[clip=true,trim=0cm 0cm 0cm 0cm,width=0.8\textwidth]
{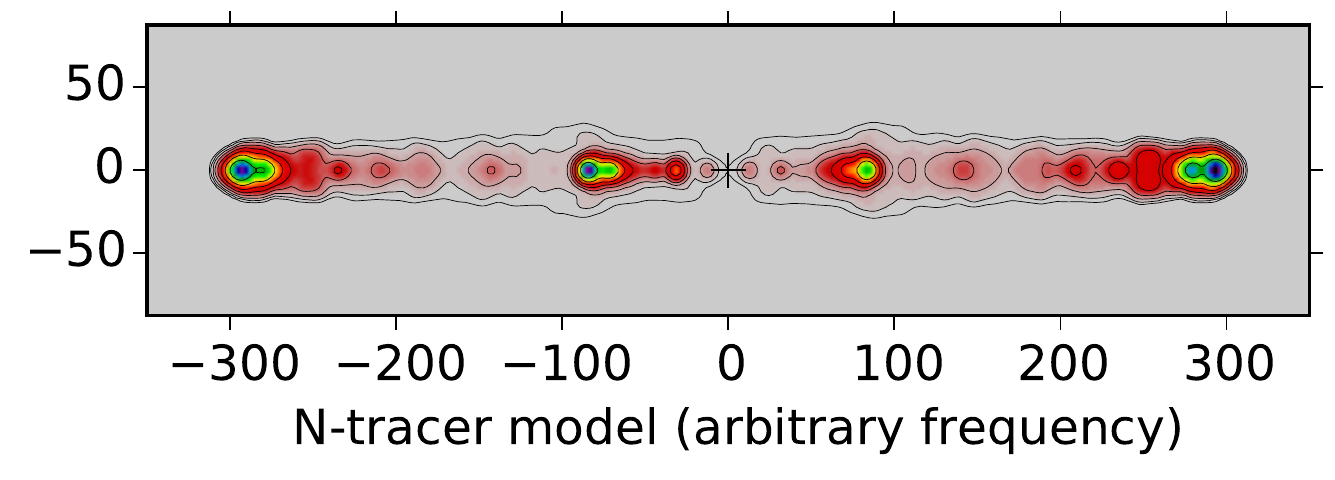}%
\put (-420,30) {\huge$\displaystyle D$}
\end{array}
$

\caption{Maps showing the influence of spectral ageing. The cut-off frequency at
the inlet of the jet is chosen to be \mbox{$\nu_{\infty} = 1$} (in arbitrary
units), whereas the
floor value of the cut-off frequency is chosen to be
\mbox{$\nu_{\infty} = 1 \times 10^{-18}$}.
Panel A: synchrotron map for
\mbox{$\nu_{\rm obs} = 1 \times 10^{-1} \: \nu_{\infty}$}.
Panel B: synchrotron map for
\mbox{$\nu_{\rm obs} = 1 \times 10^{-2} \: \nu_{\infty}$}.
Panel C: synchrotron map for
\mbox{$\nu_{\rm obs} = 1 \times 10^{-18} \: \nu_{\infty}$}.
These three images created by using the \NCooling model.
Panel D: synchrotron map for the \emph{N-tracer} synchrotron model,
with arbitrary frequency (no spectral ageing applies).
The flux levels are
\mbox{$I_{\nu} \in \{1,2,4,8,16,32,64,256\} \times I_{\nu, {\rm min}}$}, with
\mbox{$I_{\nu, {\rm min}} = I_{\nu, {\rm max}}/256$} and
\mbox{$I_{\nu, {\rm max}} = 1$}.
}
  \label{fig:5}
\end{figure*}

In Fig. \ref{fig:5} we show the effect on the radio maps of synchrotron losses
of the relativistic electrons responsible for the emission (spectral ageing,
see \ref{subsec:TheNCoolingmodel} for more details). The value of the cut-off
frequency is evaluated at each grid cell and has arbitrary units, since in our
model we just study the brightness contrasts in the synthesized images. Therefore,
we can choose \mbox{$\nu_{\infty} = 1$} at the jet-inlet. The actual break
frequency for J1835+6204 is approximately
\mbox{$2.3 \cdot 10^{4} < \nu_{\infty} < 5.9 \cdot 10^5$} MHz, with an average of
$\nu_{\infty} = 3.1 \cdot 10^{5}$ MHz (\citealt{Konar2012}).
As the material in the jets propagates towards the jet-heads, the jets expand
sideways. As a result the cut-off frequency decreases as one moves outwards from
the central engine. At the jet-head, kinetic energy from the jet material is
converted to gas pressure, resulting in a high-temperature gas that flows back in
the direction of the central engine, forming the cocoon that surrounds the jet. The
pressure gradients between the cocoon material and the undisturbed intergalactic
medium in turn also causes the cocoon to expand sideways, away from the
jet-head. As a result, the brightness of the DDRG should also decrease for
observation frequencies $\nu_{\rm obs}$ close to the cut-off frequency. This effect
is clearly visible in Fig. \ref{fig:5}: Panel A shows the DDRG for an observation
frequency \mbox{$\nu_{\rm obs} = 1\times 10^{-1} \: \nu_{\infty}$} (corresponding to
$3.1 \cdot 10^{1}$ GHz). In this case, the outer hotspots are just visible (since
the outer jets and the surrounding cocoon have expanded quite significantly),
whereas the inner jets are relatively bright (since the inner jets, embedded in the
high-pressure gas of the first jet, have not expanded that much laterally). Panel B
shows the DDRG at an observation frequency
\mbox{$\nu_{\rm obs} = 1\times 10^{-2} \: \nu_{\infty}$}. In this case, the outer
hotspots are much more prominent again, but the contours of the surrounding cocoon
are still largely absent. As the observation frequency is decreased, more and more
of the surrounding cocoon will become visible. In our study we find that the radio
structures virtually do not show any effects of synchrotron cooling for observation
frequencies \mbox{$\nu_{\rm obs} \lesssim 1\times 10^{-6} \: \nu_{\infty}$}
(corresponding to $\sim 0.3$ MHz). To make a fair comparison between the \NCooling
model (which includes the effects of spectral ageing) and the \NTracer model (which
does not include spectral ageing) we have to consider the DDRG with the \NCooling
model at sufficiently low observation frequency. These numbers imply that for our
reference radio image of J1835+6204 at 4.8 GHz, corresponding to
\mbox{$\nu_{\rm obs} \sim 1.5\times 10^{-3} \: \nu_{\infty}$} spectral ageing
effect would be notable, but would not yet drastically change the morphology of
the DDRG. The image at panel C shows the DDGR for an observation frequency of
\mbox{$\nu_{\rm obs} = 1\times 10^{-18} \: \nu_{\infty}$}, where the complete
radio structure is visible. Panel D shows the DDRG in case of the \NTracer
model (arbitrary observation frequency). Close examination of the two bottom
figures shows virtually no difference in any of the radio structures. This
strongly indicates that the \NCooling and \NTracer synchrotron models converge
at sufficiently low observation frequencies.

  \subsection{Contributions from various jet components}
  \label{subsec:JetComponents}

There are strong indicators that astrophysical jets emerging from an AGN of a radio
galaxy have a transverse radial structure (see for instance
\citealt*{Sol1989};
\citealt{Aloy2000};
\citealt{Giroletti2004};
\citealt*{Ghisellini2005};
\citealt{Gomez2008};
\citealt{Fuentes2018}
\citealt{Marti2019};
\citealt{Park2019}).
These jets are believed to consist of a low-density, high-Lorentz factor jet
spine, surrounded by a denser, slower moving jet sheath. This leads to two
distinct jet components, each of which is given a separate tracer in the
\NTracer model.
Since we would like to keep track of each of the various jet components, we
assign a total of four tracers to the \NTracer jet model:
$\theta_{\rm sp_1}$, $\theta_{\rm sh_1}$, $\theta_{\rm sp_2}$ and
$\theta_{\rm sh_2}$. The \NTracer model allows one to include the
contribution of any separate jet component to the total emissivity, in any
desired ratio. Therefore, it's possible to isolate the contribution of
single jet components, or the combination of multiple jet components to
the total emissivity.

\begin{figure*}
$
\begin{array}{c}
\includegraphics[clip=true,trim=0cm 0cm 0cm 0cm,width=0.8\textwidth]
{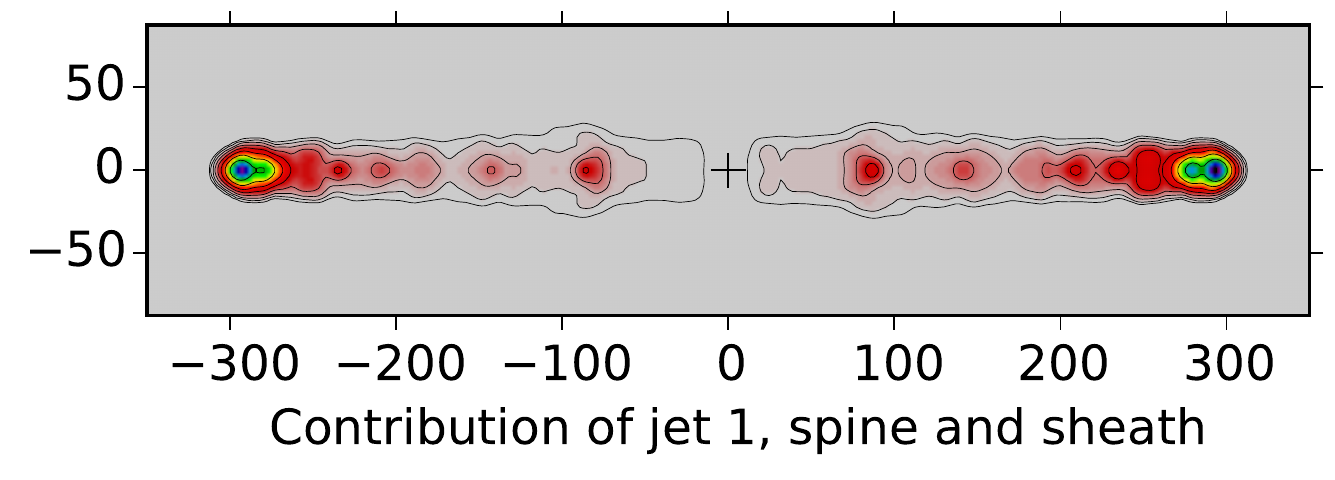}%
\put (-420,30) {\huge$\displaystyle A$}\\
\includegraphics[clip=true,trim=0cm 0cm 0cm 0cm,width=0.8\textwidth]
{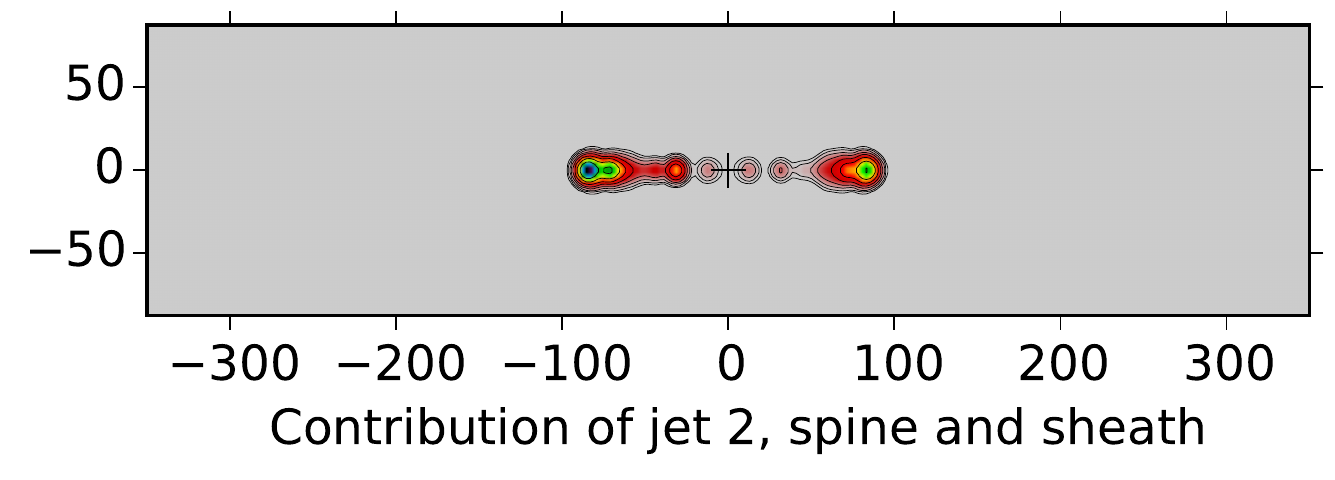}%
\put (-420,30) {\huge$\displaystyle B$}\\
\includegraphics[clip=true,trim=0cm 0cm 0cm 0cm,width=0.8\textwidth]
{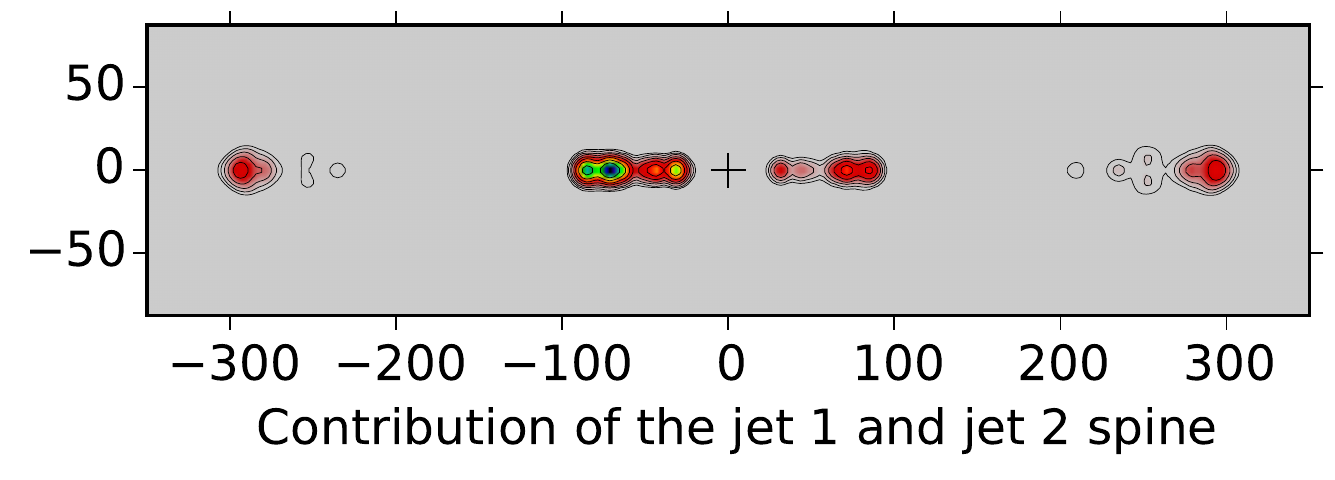}%
\put (-420,30) {\huge$\displaystyle C$}\\
\includegraphics[clip=true,trim=0cm 0cm 0cm 0cm,width=0.8\textwidth]
{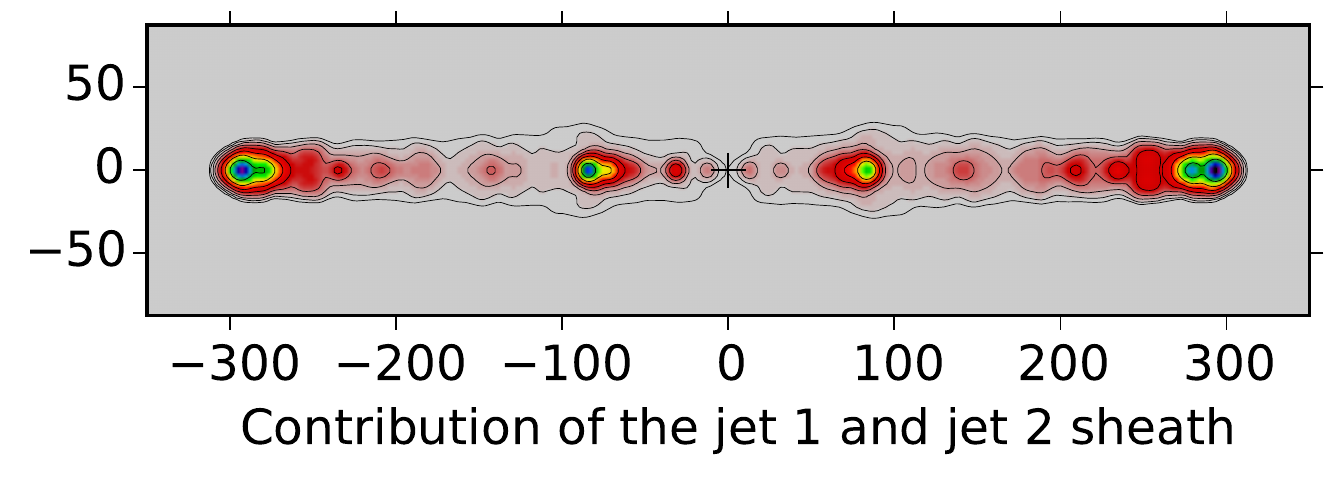}%
\put (-420,30) {\huge$\displaystyle D$}
\end{array}
$

\caption{Contributions coming from the various jet components in the case
of purely entangled magnetic fields (so $\Lambda = 1.0$). Panel A:
contribution to the total synchrotron emission coming from the initial jet
(jet 1 spine and sheath).
Panel B: contribution to the total synchrotron emission coming from the
restarted jet (jet 2 spine and sheath). Panel C: contribution to the
total synchrotron emission coming from the jet spine material (from jet 1 and 2).
Panel D: contribution coming from the jet sheath material (from jet 1 and 2).
The flux levels are
\mbox{$I_{\nu} \in \{1,2,4,8,16,32,64,256\} \times I_{\nu, {\rm min}}$}, with
\mbox{$I_{\nu, {\rm min}} = I_{\nu, {\rm max}}/256$} and
\mbox{$I_{\nu, {\rm max}} = 1$}.
}
  \label{fig:6}
\end{figure*}

Fig. \ref{fig:6} shows the jets at an age of 16.6 Myr (as before), in case
of the \NTracer model. We've assumed a purely entangled magnetic
field configuration (so $\Lambda = 1$). We show the contribution of
the initial jets to the total emission ($\theta_{\rm sp_1}$ and
$\theta_{\rm sh_1}$) in panel A; the contribution of the restarted
jets to the total emission ($\theta_{\rm sp_2}$ and $\theta_{\rm sh_2}$) in
panel B; the contribution of the jet spine material to the total emission
($\theta_{\rm sp_1}$ and $\theta_{\rm sp_2}$) in panel C; and the
contribution of the jet sheath material to the total emission
($\theta_{\rm sh_1}$ and $\theta_{\rm sh_2}$) in panel D.
%

%
%

In Fig. \ref{fig:6} A, the contribution of the initial jets
is isolated. The two outer hotspots are clearly visible, and the inner jets
are absent. Moreover, the structure of the cocoon and radio lobes shows little
differences with that of the complete image that includes the contribution of
both the initial jets and the restarted jets. Surprisingly, we see that
there are two very clear `warm spots' just ahead of where the termination shock
of the second jets should be. Apparently, left-over jet material from the first
jet eruption is locally compressed so much by the head of the restarted jet that
it significantly contributes to the total emissivity in the synthesized map.

In Fig. \ref{fig:6} B, the contribution of the restarted jets
is isolated. Since these restarted jets have not propagated that far along the axis
of the system, the associated radio features are relatively compact,
much thinner, and the restarted jets are much more stable. This is reflected in the
image: the cocoon is localised near the jet axis and more of the jet itself is
visible.

Fig. \ref{fig:6} C shows the contribution of the jet spine material of both the
initial jets and the restarted jets to the total emission of the synthesized maps.
First of all, the radio features of the spine of the restarted jet (shown here
together with the spine of the initial jet) seem somewhat thinner than panel B,
where the full (spine+sheath) restarted jet is shown. This indicates that the
restarted jet maintains its radial structure. For the initial jet, most of the lobe
material is not seen in this map, except for a small contribution to the total
emission in the outermost hotspots of the initial jets. This implies that almost
all of the radio structures of the initial jets is due to the jet sheath material.

Looking at Fig. \ref{fig:6} D, this suspicion is confirmed. There, the image shows
the contribution of the jet sheath material of the initial jets and of the
restarted jets. This image shows great similarities with the image of the total
emissivity, in fact, the radio structures of the outer radio lobes is almost
identical. It is clear that the brightness contrasts of the inner jets differ
from the brightness contrasts in the map of the total emissivity: the inner jets
seem to be less continuous and the inner structures are less bright. 

  \subsection{Effects of viewing angle}
  \label{subsec:ViewingAngle}

\begin{figure*}
$
\begin{array}{c}
\includegraphics[clip=true,trim=0cm 0cm 0cm 0cm,width=0.8\textwidth]
{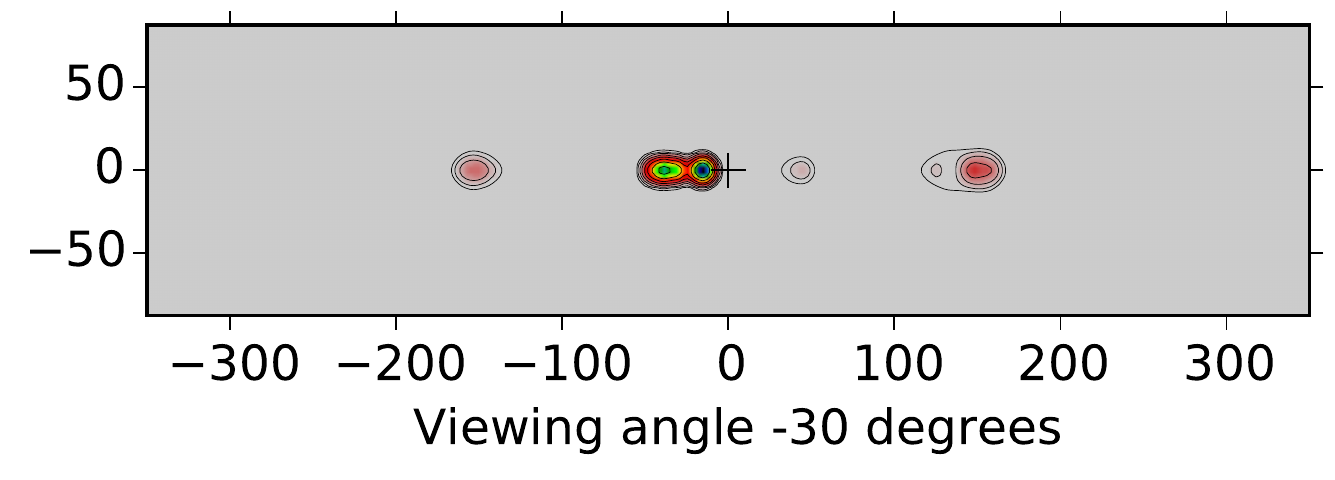}%
\put (-420,30) {\huge$\displaystyle A$}\\
\includegraphics[clip=true,trim=0cm 0cm 0cm 0cm,width=0.8\textwidth]
{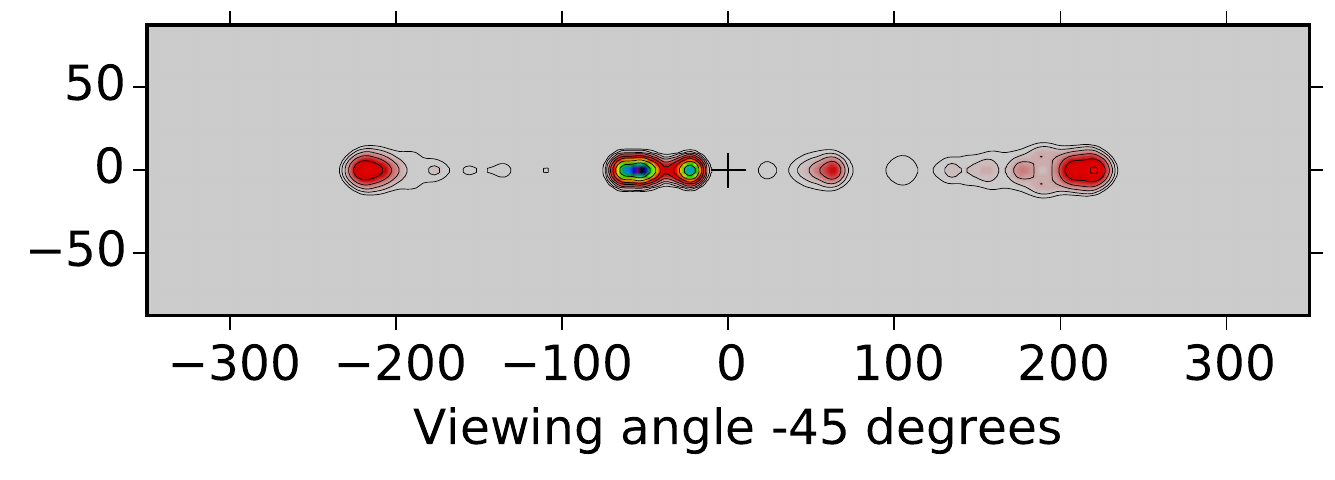}%
\put (-420,30) {\huge$\displaystyle B$}\\
\includegraphics[clip=true,trim=0cm 0cm 0cm 0cm,width=0.8\textwidth]
{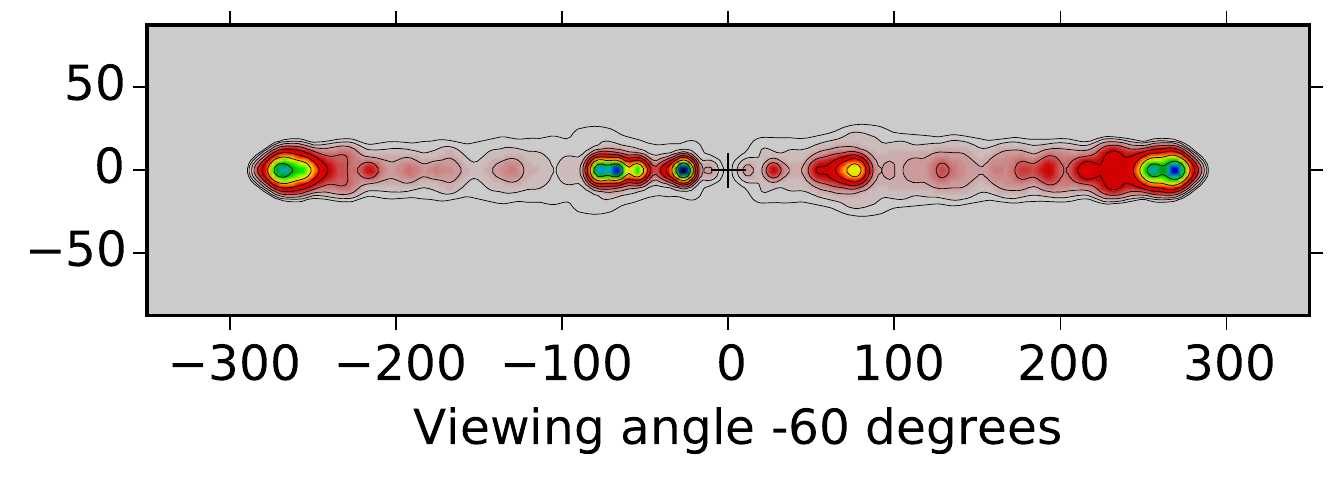}%
\put (-420,30) {\huge$\displaystyle C$}\\
\includegraphics[clip=true,trim=0cm 0cm 0cm 0cm,width=0.8\textwidth]
{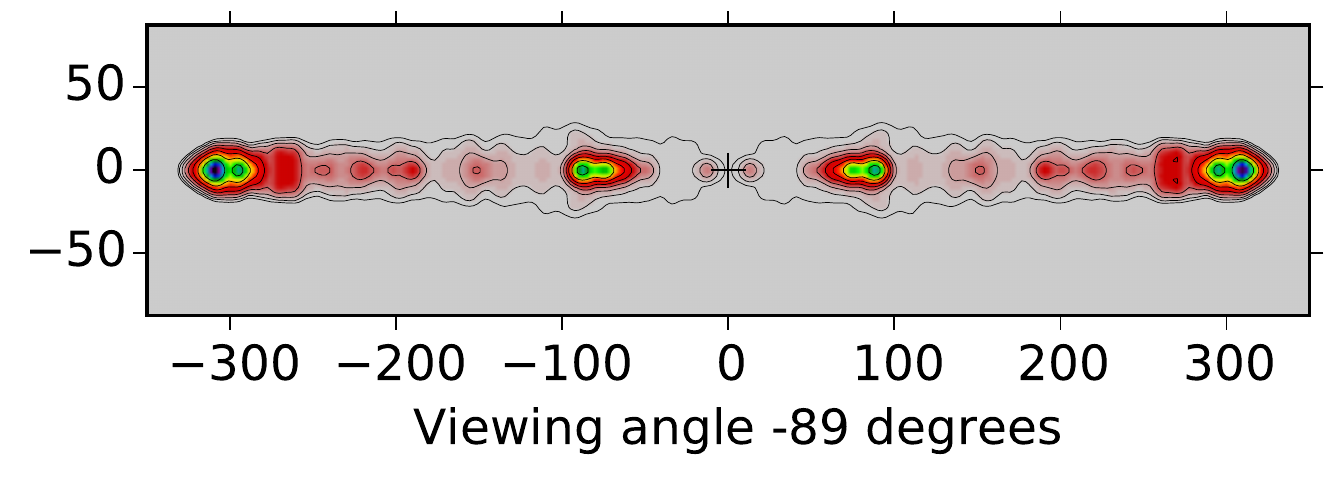}%
\put (-420,30) {\huge$\displaystyle D$}
\end{array}
$

\caption{Influence of viewing angle, Doppler factor in the case of purely
entangled magnetic fields (so $\Lambda = 1.0$). Viewing angles from Panel A to D:
-30$^\circ$, -45$^\circ$,-60$^\circ$, -89$^\circ$.
In this convention, a viewing angle between -90 degrees and 0 degrees correspond
to the lower part of the jet pointing towards the observer. For 0 degrees, the
jets are aligned with the line of sight and for 90 degrees, the jets are
perpendicular to the line of sight.
The flux levels are
\mbox{$I_{\nu} \in \{1,2,4,8,16,32,64,256\} \times I_{\nu, {\rm min}}$}, with
\mbox{$I_{\nu, {\rm min}} = I_{\nu, {\rm max}}/256$} and
\mbox{$I_{\nu, {\rm max}} = 1$}.
}
  \label{fig:7}
\end{figure*}

In Fig. \ref{fig:7} the effect of a changing viewing angle is shown. Panel A shows
a viewing angle of \mbox{$\vartheta = - 30^{\circ}$}; Panel B shows a viewing angle
of \mbox{$\vartheta = - 45^{\circ}$}; panel C shows a viewing angle of
\mbox{$\vartheta = - 60^{\circ}$}; and panel D shows a viewing angle of
\mbox{$\vartheta = - 89^{\circ}$} (so almost face on, but still slightly tilted to
break the symmetry between the NW and SE jets).

The observed size of the DDRG $D_{\rm obs}$ is determined by the actual size of
the source, $D_{\rm DDRG}$, multiplied by $\sin{\vartheta}$, so the smaller the
viewing angle, the smaller the observed size of the source. The viewing angle
has an effect on two quantities, namely [1] the Doppler factor and [2] the
perpendicular magnetic field component of an ordered magnetic field configuration.
Below we only discuss the effects of the viewing angle for jets with an entangled
field configuration. This means that a change in the features is completely
attributed to a change in the Doppler factor, assuming that the coherence length
of the turbulent magnetic field is much smaller than the scale corresponding to
the resolution adopted in these images.

The observed synchrotron emissivity at a certain frequency $\nu$ is very
sensitive to variations in the velocity and viewing angle through the Doppler
factor, as given by \equref{eq:jnuGeneral}. Since the inner jets propagate through
a much more dilute medium than the outer jets, their jet-head advance speed (and
therefore also advance speed of the hotspots) is much larger. In our models
we find for the outer jet-head $\beta_{\rm hd_1} \approx 0.045$, while for the
inner jets we find $\beta_{\rm hd_2} \approx 0.7$, roughly a factor of 16 times
larger (see $\SWb$). These values compare quite well to those inferred from
observations, such as J1835+6204 (see for example \citealt{Konar2012}). Based on
these values, we find that the Doppler factor of the approaching inner jet is
larger than that of the approaching outer jet for viewing angles
$-65^{\circ} < \vartheta < 0^{\circ}$. For $-20^{\circ} < \vartheta < 0^{\circ}$,
the ratio of the Doppler factor of the inner jet to that of the outer jet is even
more than a factor of $\mathcal{D}_{\rm in}/\mathcal{D}_{\rm out} > 2$. As a
consequence the observed synchrotron emission, which scales as $D^{2+\alpha}$,
can be a factor of 5--10 times as high as for the outer jets. As a result, when
the source is tilted toward the observer so that the viewing angle becomes
smaller, the synchrotron emission of the approaching inner jet is able to outshine
its corresponding outer jet. This can best be seen in panel A where
\mbox{$\vartheta = - 30^{\circ}$}. There the SE2 (inner) jet is significantly
Doppler boosted, while the NW2 (inner) jet is so Doppler dimmed that it is almost
no longer visible. Moreover, both the (outer) NW1 and SE1 jets are very weak
compared to, and are outshined by, the SE2 jet.

The brightness contrast between the outer NW1 and SE1 jets is not that big, which
implies that the Doppler (de-)boosting is not strong for these jets. In fact,
close examination shows that the NW1 hotspot is actually brighter than the
SE1 jet. This can only be explained by the effect of the Doppler factor associated
with the back-flowing material in the cocoon of the receding jet. 
This opens the possibility that a telescope with a limited dynamic range does not
observe the outer hotspots or radio features of a DDRG at all if the viewing
angle of that source is too small.

The general effects of the Doppler (de-)boosting on the morphology of the DDRG
is observed at a wide range of viewing angles. Its effect becomes less strong
for viewing angles near $\pm 90^{\circ}$.
At \mbox{$\vartheta = -45^{\circ}$} the outer and inner jets are still completely
disconnected, but at \mbox{$\vartheta = - 60^{\circ}$} the outer NW1 contour
encloses the NW2 jet. With large viewing angles both inner jets are enclosed by
their outer counterparts, as can be seen in panel D.

  \subsection{Comparing different stages of evolution}
  \label{subsec:StagesOfEvultion}

\begin{figure*}
$
\begin{array}{c}
\includegraphics[clip=true,trim=0cm 0cm 0cm 0cm,width=0.8\textwidth]
{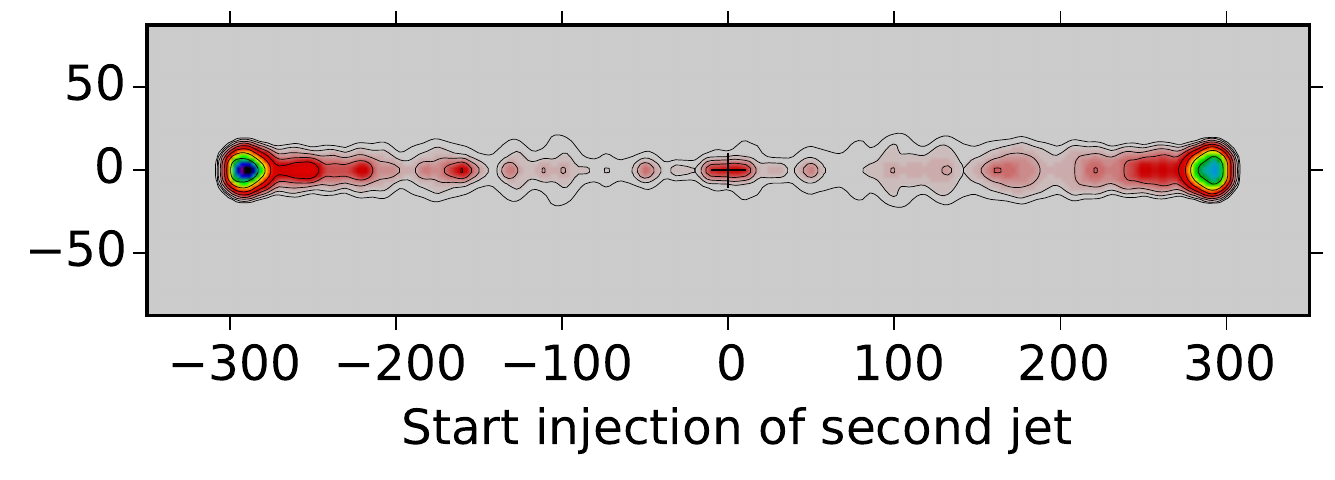}%
\put (-420,30) {\huge$\displaystyle A$}\\
\includegraphics[clip=true,trim=0cm 0cm 0cm 0cm,width=0.8\textwidth]
{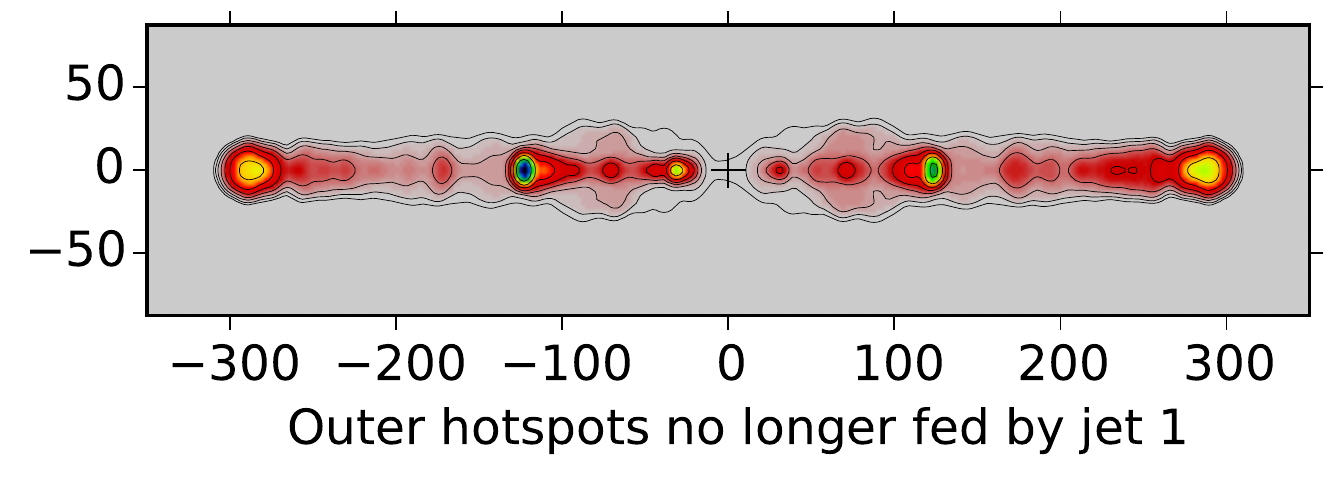}%
\put (-420,30) {\huge$\displaystyle B$}\\
\includegraphics[clip=true,trim=0cm 0cm 0cm 0cm,width=0.8\textwidth]
{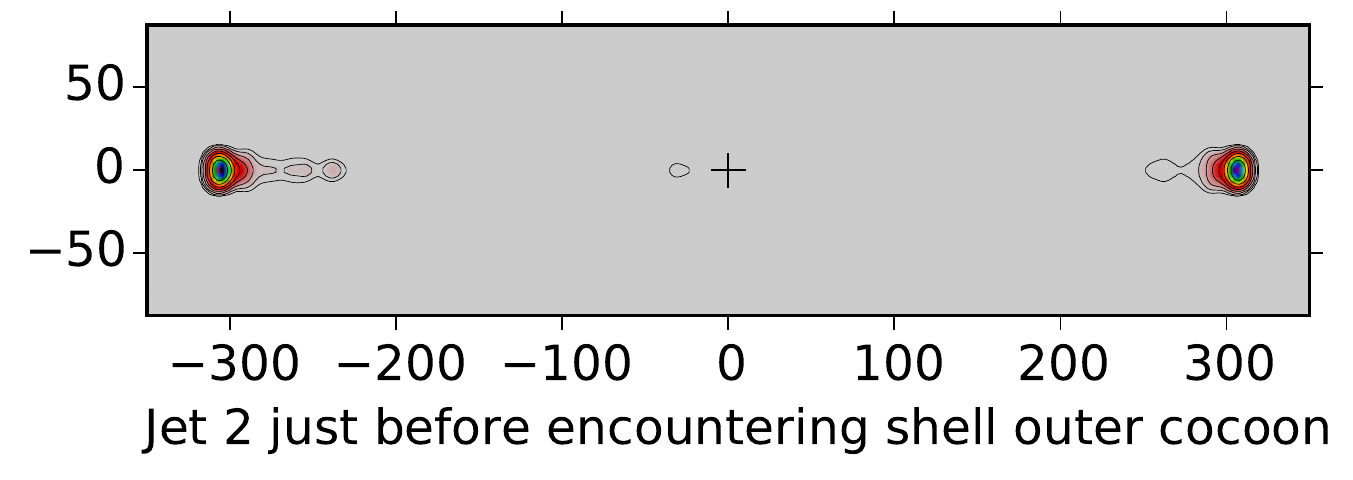}%
\put (-420,30) {\huge$\displaystyle C$}\\
\includegraphics[clip=true,trim=0cm 0cm 0cm 0cm,width=0.8\textwidth]
{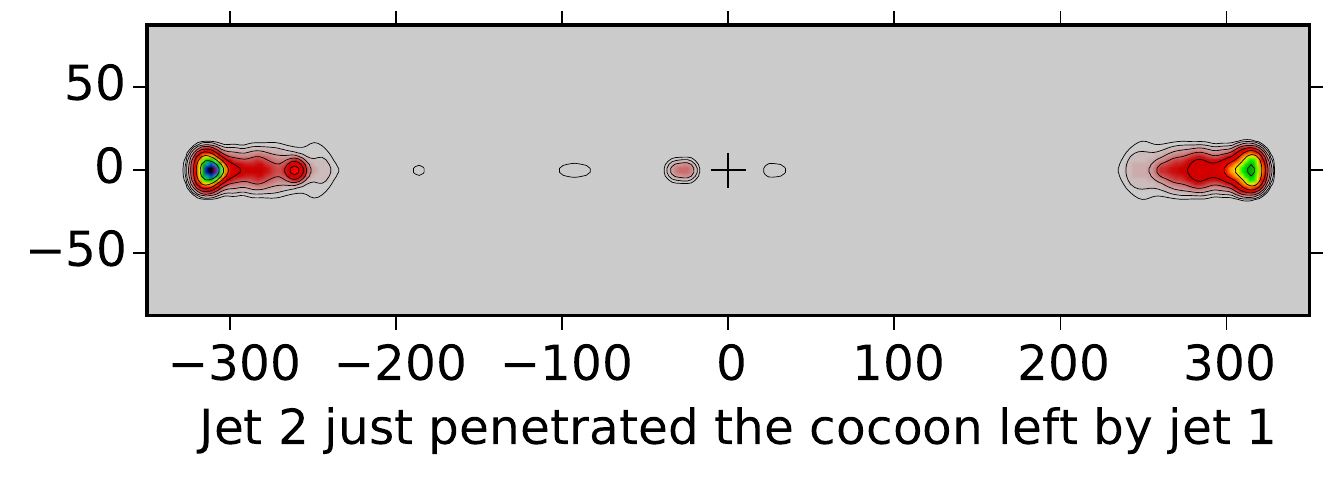}%
\put (-420,30) {\huge$\displaystyle D$}
\end{array}
$

\caption{The isochoric jet model A2 shown at various stages during the evolution.
Panel A: first jet at 16.0 Myr, just before the second jet is turned on.
Panel B: 16.65 Myr, shortly after the outer hotspots are no longer fed
by the first jets.
Panel C: second jets at 18.0 Myr, just before the second jets will encounter
the shell of the outer cocoon.
Panel D: second jets at 18.7 Myr, shortly after the second jets have broken
out of the outer cocoon.
The flux levels are
\mbox{$I_{\nu} \in \{1,2,4,8,16,32,64,256\} \times I_{\nu, {\rm min}}$}, with
\mbox{$I_{\nu, {\rm min}} = I_{\nu, {\rm max}}/256$} and
\mbox{$I_{\nu, {\rm max}} = 1$}.
}
  \label{fig:8}
\end{figure*}

The jet-head advance speed of the restarted jets is much higher than that
of the initial jets. In fact, the restarted jets only reside within the outer
cocoon for a small fraction of time, compared to the age of the source.
In Fig. \ref{fig:8}, we show a few important phases of the evolution of a DDRG.
Panel A shows the DDRG at a time of 16.0 Myr, where the initial jets
have been switched off for approximately $7 \times 10^5$ years. The hotspots
are still being fed by the initial jets that have not yet dissipated. This
observation was also made in the previous paper $\SWb$, where a clear tail of
jet material with Lorentz factor well above 1 could be seen. Even though
the tails of the jets have moved outwards compared to the central engine, radio
features are still visible close to the central engine.
At 16.0 Myr, the restarted jets are injected and the inner jets propagate
through the cocoon of the initial jets, while the leftover initial jets still
feed the outer hotspots. This phase continues until the tails of the initial
jets reach the outer hotspots, when a new phase starts:
In panel B, the DDRG is shown at a time of 16.65 Myr, just after the
initial jets have dissipated and the outer hotspots are no longer fed by jet
material. The outer cocoon is still clearly visible, but its brightness
weakens as the system expands adiabatically.
At a certain point the restarted jets will reach the edge of the outer cocoon.
This boundary of the outer cocoon is a relatively thick shell (whose density
can be a factor $> 4$ more than that of the undisturbed intergalactic medium).
This means that as soon as the restarted jets
start to interact with the outer
cocoon a very strong termination shock forms. At this point, all other radio
features that are contained within the cocoon are outshined by the newly formed
termination shocks and only two very bright hotspots will be visible. This
situation is shown in panel C of Fig. \ref{fig:8} at a time of
18.0 Myr.
Because the shell of the (outer) cocoon is so dense, it takes the inner jets
quite a long time to completely penetrate them, in our simulations approximately
$7 \times 10^5$ years. The back-flow of jet material in this phase is also
relatively strong. Panel D shows the system at 18.7 Myr, just after the
restarted jets have just penetrated the outer cocoon.
%

\clearpage
%

\clearpage

    \subsection{Differential light travel times}
    \label{subsec:DifferentialLTT} 

The synthesized maps shown here do not take account of (differential) light travel
time effects between the source and a distant observer, which affect the appearance
of the source for such an observer. When the jets from a radio galaxy have a viewing
angle \mbox{$\vartheta \ne 90^{\circ}$}, the distance from the observer to the
advancing jet-head ($L_{\rm a}$) is less then the distance from the observer to
the receding jet-head ($L_{\rm r}$). The distance from Earth to an extra-galactic
radio galaxy ($D_{\rm s}$) is always much larger than the jet length
($D_{\rm jt}$). Then the difference in light travel time to the observer
for the advancing and the receding jet-head equals:

\be
\Delta t \; \propto \; \frac{L_{\rm r} - L_{\rm a}}{c} \approx
\frac{2D_{\rm jt} \cos{\vartheta}}{c} \; ,
\ee
where we have assumed that the advancing jet and the receding jet have equal
length. This means that at a given moment of observation the recorded image
of the jets becomes ``progressively younger" if one moves from the receding
hotspot towards the approaching hotspot. This is not included in our model. When
viewing the synthesized synchrotron maps in this paper, one should bare in mind
that the receding jet will appear younger (and therefore shorter) in reality. Almost
all the synthesized synchrotron maps that we show assume a viewing angle of
\mbox{$\vartheta = -71^{\circ}$}. The distance from the outer jet-heads to the AGN
have an approximate size of \mbox{$D_{\rm jt_1} \approx 300$ kpc}, while for the inner
jet-heads we find \mbox{$D_{\rm jt_2} \approx 90$ kpc}. Then the maximum time
difference ($\Delta t$) per age of the AGN ($t_{\rm s}$) for the outer- and
inner jet-heads respectively equals:

\be
\frac{\Delta t}{t_{\rm s}} \; \approx \;
\frac{2 \: \cos{(-71^{\circ})}\: D_{\rm jt_1}}{c \: t_{\rm s_1}} \approx 4%
      \mathrm{\; per\; cent} \; ,
\ee

\be
\frac{\Delta t}{t_{\rm s}} \; \approx \;
\frac{2 \: \cos{(-71^{\circ})}\: D_{\rm jt_2}}{c \: t_{\rm s_2}} \approx 32%
\mathrm{\; per\; cent} \; ,
\ee
so that our neglect of the differential light travel effects for the inner
jets should in reality be improved upon. We refer the reader to $\SWb$, Fig. 3
for detailed information on the position of the jet-head of the first and the
second jet, corresponding to the jet-length $D_{\rm jt_1}$ and $D_{\rm jt_2}$.

\section{Conclusions}
\label{sec:Conclusions}

In this paper we have used the simulations of three different jet models in
order to synthesize synchrotron radiation images at radio frequencies for
double-double radio galaxies (DDRGs). As a reference source, we have chosen the
DDRG J1835+6204 to compare our results with. The jet models are the homogeneous jet
$H2$; the isothermal jet $I2$; and the (piecewise) isochoric jet $A2$. In order
to calculate the emissivity at frequency $\nu$, we calculated the Doppler factor;
the (space-time dependent part of the) relativistic particle distribution
$\mathcal{N}$; and the magnitude of the magnetic field component perpendicular
to the line of sight $B_{\perp}$. We performed pure hydrodynamical simulations,
and therefore approximated the (ordered, entangled, or mixed) magnetic fields.
While this clearly is an approximation, it allowed us to
explore the effect of field line geometry on the radio images. In this work we
have studied the effects on the brightness contrasts of the synthesized synchrotron
images of:
\begin{itemize}
\item[1] the radial profiles of the three jet models $H2$, $I2$ and $A2$;
\item[2] the various magnetic field configurations;
\item[3] spectral ageing from synchrotron cooling;
\item[4] contributions from jet components
($\theta_{\rm sp_1}$, $\theta_{\rm sh_1}$,
$\theta_{\rm sp_2}$, $\theta_{\rm sh_2}$);
\item[5] the viewing angle of the jets;
\item[6] the various epochs (phases) of the evolution of a DDRG.
\end{itemize}
We find the following conclusions:

\begin {itemize}
\item[(i)] In all cases (where entangled magnetic fields are involved) the
synthesized synchrotron images show two clear (pairs of) hotspots at 8 kpc
resolution (corresponding to a VLA radio observation at 4.8 GHz, and average
beam size of 1.4''): The outer hotspots from the first (initial) jets,
NW1 and SE1, and the inner hotspots from the second (restarted) jets,
NW2 and SE2, and therefore satisfying the condition for a DDRG;

\item[(ii)] In all cases
the expected effects of Doppler boosting/dimming are confirmed:
we find that for jets that are approaching the observer,
the associated hotspots and other radio features along its jet axis are Doppler
boosted (brightened), whereas for the jets that are receding, these same radio
features are Doppler deboosted (dimmed). Notably, the opposite behaviour in
the back-flowing material can also clearly be seen;

\item[(iii)] For the observed DDRG J1835+6204 at 8 kpc resolution and
4.8 GHz, we find the closest resemblance to the observations with the
(piecewise) isochoric jet model $A2$. That model generates synthesized
synchrotron images consistent with J1835+6204, due to its relatively low radial
jet integrity;

\item[(iv)] We find that the synchrotron radiation that is generated by the 
ordered magnetic fields appears as a small number of knots along
the jet axis, in the direct vicinity of internal shocks, or just before the
Mach disc of the jet-heads. When assuming a mix of entangled magnetic fields
and ordered magnetic fields, we notice a shift in brightness contrasts. When the
magnetic pressure of the ordered fields become significant ($\gtrsim 50$ per cent
of the total magnetic pressure), details of the cocoon start to fade in the
synthesized images. We get the best resemblance with the observation of J1835+6204
when the magnetic pressure of the ordered fields is small compared to the total
magnetic pressure ($\lesssim 30$ per cent, so $\Lambda \gtrsim 0.7$);

\item[(v)] The effect of spectral ageing by synchrotron cooling becomes
apparent at observation frequencies
\mbox{$\nu_{\rm obs}/\nu_{\rm\infty,0} \gtrsim 10^{-6}$}, and becomes
significant at frequencies \mbox{$\nu_{\rm obs}/\nu_{\rm\infty,0} \gtrsim 10^{-3}$},
where \mbox{$\nu_{\rm\infty,0} \equiv 1$} denotes the cut-off frequency of the
relativistic particle population at jet inlet. For J1835+6204 this cut-off
frequency is approximately $\nu_{\infty} \approx 3.1 \cdot 10^{5}$ MHz.
The most notable effect of spectral ageing in the synthesized synchrotron
images is the fading of the cocoon. At relatively very high observation
frequencies \mbox{$\nu_{\rm obs}/\nu_{\rm\infty,0} \gtrsim 10^{-1}$} the effects
of spectral ageing becomes so strong that also the outer hotspots of the jets
NW1 and SE1 will vanish in the synchrotron maps. Taking the cut-off frequency for
J1835+6204, this would then occur at an observation frequency of
\mbox{$\nu_{\rm obs} \sim 3 \cdot 10^1$ GHz}.

\item[(vi)] We are able to separate the contributions in synchrotron radiation
coming from the different jet components (the spine material, or the sheath
material from the first jet, and equivalently for
the second jet). We find that the first jets create a wide cocoon. The
second jets create a very thin cocoon. The bow shock of disturbed
material within the outer cocoon, which is pushed by the second jets is also
quite bright, so that the inner hotspots also partially lightened up by the
left-over material from the first jets. Finally we find that the
contribution from jet spine material remains very close to the jet axis, while
the contribution from the jet sheath material comes from both the jet and the
surrounding cocoon.

\item[(vii)] The viewing angle has a strong effect on the observed size of the
source, as well as the Doppler boosting/dimming of the jets. The Doppler
effect is strongest for the inner jets, due to the fact that the jet-head
advance speed of the inner jets ($\sim 0.7$c) is higher than that of
the remaining outer jets ($\sim 0.045$c).
When the viewing angle $\vartheta \sim -30^{\circ}$, the approaching inner jet
is brightest, while the receding inner jet has virtually faded completely. Also
the outer jets have almost faded completely. For even smaller viewing angles all
that remains in the synthesized image will be the approaching inner jet:
A DDRG structure would no longer be recognisable.

\item[(viii)] The time it takes for the first jet to fade after the central
engine has turned off is small compared to the life time of the radio source.
Therefore, the chance of detecting an episodic AGN jet radio source
as an actual DDRG is small: the second jets must be turned on again well before
the first jets have completely faded. Within a few hundred thousand years the
radio morphology of the source would look completely different. We find that
when the restarted jets start to penetrate the cocoon that was left behind by
the first jets nearly all that is visible are the termination shocks at the
jet-heads. It is at this point that a strong back flow begins to form again,
and more of the cocoon structure will arise again.
\end {itemize}

  \subsection{Outlook}
  \label{subsec:Outlook}

From the results we've been able to derive contributions from magnetic field
components for the entangled, the azimuthal, the poloidal and the helical, and
the mixed magnetic field configurations. In order to do so, we had to make the
necessary approximations and assumptions. Adding actual magnetic fields to this
model (switch to a full MHD module) would be a valuable contribution.
Such SRMHD simulations would also allow for a more self-consistent
quantification of the measured polarization. A few more aspects are mentioned
in this outlook, to be considered for future work.

\subsubsection{Inverse-Compton losses}\label{subsubsec:IC}

When discussing spectral ageing, we ignored the potential influence of
Inverse-Compton losses, which is likely a significant contributor to radiative
losses. \cite{Hardcastle2018} discusses a semi-analytical model for the evolution
of powerful radio galaxies, where trends between jet power and luminosity are
reproduced. Inverse Compton losses arise from scattering of the Cosmic Microwave
Background, whose radiation field energy density is about
$U_{\mathrm{rad}}\approx 4.17 \times 10^{-13} (1+z)^4 \,
{\mathrm{erg}}/{\mathrm{cm}}^3$ in intergalactic space at redshift $z$. In
practice, this would imply that our energy loss formulae must replace the
magnetic field with an effective field:
\be
\frac{B_{\mathrm{eff}}^2}{8\pi} = \frac{B^2}{8\pi}+U_{\mathrm{rad}} \,.
\ee
At the same time, the synchrotron emission must still employ the actual field
pressure ${B^2}/{8\pi}$. Since this in practice introduces yet another degree of
freedom when generating radio maps from our SRHD simulations, we did not
incorporate this effect here. In future SRMHD runs, where the ${B^2}/{8\pi}$ is
self-consistently computed, these losses should be considered too.

    \subsubsection{Dimensionality of the simulations}
    \label{subsubsec:Dimensionality} 

These simulations have been performed in a 2.5D (cylindrical symmetric) setting.
Many observed features give a good notion on the processes at hand. However, in a
full 3D setting the jet flow, as well as the back-flowing material are able to
effectively propagate in one extra dimension. This causes a less symmetrical flow
and allows for more instabilities to arise. For example, eddies (vortices) that
form at the jet-head in the back-flowing material in the 2.5D case all have
approximately the same size and show similar flowing behaviour.
They propagate downwards
relatively steady along the jet-axis, away from the jet-head. In a 3D case, these
eddies will most likely have different shapes and not form with the
same degree of symmetry as that imposed in a 2.5D setting. Moreover, it is to be
expected that the jet axis itself will begin to wobble under the influences of
instabilities, instead of being a perfectly straight line. This will inevitably
have an effect on the jet-cocoon interaction, the related jet pinches and internal
shocks, which in turn all have an effect on the related observed radio features.
Therefore, the results should be interpreted with some caution. The advantage of
performing 2.5D simulations compared to 3D simulations is that we are able to
resolve radio structures in much more detail, while being able to let the jets
evolve on a very large scale.

    \subsubsection{Renewed injection of non-thermal particles}
    \label{subsubsec:RenewedInjection} 

In this work we have not involved the injection of a fresh non-thermal particle
population at internal shocks, or at the termination shock of the jets. If these
fresh injections were to be implemented, the brightness contrasts within the
synthesized images would of course change. The effect of injecting fresh
relativistic particles will, however, become most apparent at those frequencies
were spectral ageing has already played a significant role. A realistic
implementation of the injection of fresh populations of relativistic particles
would require an observational study of particle populations near internal shocks
and the termination shock.

\section*{Acknowledgements}
This research is funded by the {\it Nederlandse Onderzoekschool Voor Astronomie}
(NOVA).
RK acknowledges a joint FWO-NSFC grant G0E9619N and from Internal Funds KU Leuven,
project C14/19/089 TRACESpace. We thank the anonymous referee for useful feedback
and suggestions that improved our manuscript.

\section*{Data availability}
The data underlying this article will be shared on reasonable request to the
corresponding author.

\footnotesize{
\bibliographystyle{./mn2e} 
\bibliography{References}
}

\bsp

\appendix

\section{}
  \label{sec:RelativisticParticles}

  \subsection{Advecting a populations of relativistic leptons: adiabatic limit}
    \label{subsec:TheNTracermodel}

In this Appendix we briefly consider the transport equation for relativistic leptons
(electrons and/or positrons) in a relativistic flow. The basic assumption is that
leptons are advected passively, with negligible diffusion with respect to the flow.
This implies strong collisional coupling between the bulk flow and the population of
leptons, mediated by scattering due to low-frequency MHD waves. In this situation the
leptons change their energy: they lose energy due to radiation losses (mostly
synchrotron losses) and expansion losses, and can gain energy if the flow is
compressed. Notation-wise we follow the main paper:
\mbox{$\epsilon' = \sqrt{|\bm{p}'|^2+m_{\rm e}^2}$} is the particle energy in the
local fluid rest frame (FRF) (i.e. comoving energy), in units where \mbox{$c = 1$}
(which we will assume in the appendices from this point on), and where $m_{\rm e}$
is the electron rest-mass. Quite generally we will employ primes to denote quantities
measured in the local fluid rest frame. We employ (following \citealt{Achterberg2018a})
mixed phase-space variables, where particle momentum is measured in the FRF, but
space-time position $x^{\mu} = (\bm{r} \: , \: t)$ is measured in the laboratory frame
(observer\rq{s} frame). Neglecting the effects of diffusion with respect to the flow
and viscous heating of the leptons the proper particle number density per unit comoving
energy in mixed variables,
\be
{\cal N}_{\epsprime}(\bm{r} \: ,\: t) \: = \:
{\cal N}(\bm{r} \: , \: t \: , \: \epsprime) \: \equiv \:
\frac{{\rm d} \nprime}{{\rm d} \epsprime}
\ee
satisfies, neglecting radiation losses for now as well as the possible injection of
fresh leptons,
\be
\label{simpletransp}
	\frac{\partial}{\partial x^{\mu}} \left( U^{\mu} \: {\cal N}_{\epsprime} \right) -
	\frac{\partial}{\partial \epsprime} \left[ \frac{\epsprime}{3} \:	
	\left( \nabla \bdot U \right) \: {\cal N}_{\epsprime} \right] \: = \: 0 \; .
\ee
We employ a covariant
notation for the four-divergence of the plasma flow:
\be
	\nabla \bdot U \: \equiv \: \frac{\partial U^{\mu}}{\partial x^{\mu}}
	\: = \: \frac{\partial \gamma}{\partial t} + \grad \bdot \bm{U} \; ,
\ee
with four-velocity $U^{\mu} = (\gamma \: , \: \bm{U})$ where $\bm{U} = \gamma \: \bm{V}$
and $\gamma = 1/\sqrt{1 - |\bm{V}|^2}$ the bulk flow Lorentz factor. 
The second term on the left-hand side of (\ref{simpletransp}) describes the energy change
due to expansion losses (or compression gains), c.f. \citet{Webb1985} and
\citet{Achterberg2018a}, Eqn. 8.95:
\be
\label{Echange}
	\frac{{\rm d} \epsprime}{{\rm d} \tau} \: =\:
       -\frac{\epsprime}{3} \: \left( \nabla \bdot U \right) \: =\:
       -\frac{\epsprime}{3} \:
	\left(\frac{\partial \gamma}{\partial t} + \grad \bdot \bm{U} \right) \; .
\ee
Here
\be
	\frac{{\rm d}}{{\rm d} \tau} \:=\: U^{\mu} \: \frac{\partial}{\partial x^{\mu}}
	\:=\: \gamma \: \left( \frac{\partial}{\partial t} + (\bm{V} \bdot \grad) \right)
	\: \equiv \: \gamma \: \frac{{\rm d}}{{\rm d}t} 
\ee
\halfskip \noindent
is the covariant comoving derivative in the bulk flow. 
Mass conservation in the bulk flow with lab frame density $\rho = \gamma \rho'$ reads
\be
	\frac{\partial\rho}{\partial t} + \grad \bdot(\rho\:\bm{V}) \: = \:
	\frac{\partial(\gamma\:\rho')}{\partial t} + \grad \bdot(\rho'\:\bm{U})\: = \: 0\;,
\label{eq:RelMassConservation}
\ee
or equivalently
\be
\label{altdens}
	\frac{1}{\rhoprime} \: \frac{{\rm d} \rhoprime}{{\rm d}t} \: = \:
	-\left(\frac{1}{\gamma} \: \frac{{\rm d} \gamma}{{\rm d}t} + \grad \bdot \bm{V}
	\right) \; .
\ee
For a power-law distribution in \mbox{comoving energy} of
the form
\be
	{\cal N}_{\epsprime} \: = \:
	{\cal N}(\bm{r} \: , \: t) \:  \left(\epsprime \right)^{-s} \; ,
\ee
the transport equation can be written as
\be
\label{constr}
	\gamma \: \frac{{\rm d} {\cal N}}{{\rm d} t} + \left(\frac{s+2}{3} \right) \:
	\left(\frac{\partial \gamma}{\partial t} + \grad \bdot \bm{U} \right){\cal N}
	\: = \: 0 \; .
\ee
Then Eqn. (\ref{altdens}) for the proper plasma density $\rhoprime$ implies:
\be
	\mathcal{N}\: =\: \mathcal{N}_0 \:
	\left( \frac{\rhoprime}{\rhoprime_0} \right)^{(s+2)/3} \;
	\propto \; \left(\rho' \right)^{(s+2)/3} \; ,
\label{eq:N-tracer}
\ee
with $\mathcal{N}_0({\bm r}\:,\:t)$ and $\rhoprime_0({\bm r}\:,\:t)$ corresponding
to the values at position $\bm{r}_0$ at time $t_0$. This is a good approximation
for those energies well below the energy where the synchrotron break in the
spectrum occurs and synchrotron losses need to be taken into account.
\nskip
For ultra-relativistic particles the energy density in the plasma rest-frame,
which we denote simply by ${\cal W}$, equals
\be
{\cal W}(\bm{r} \: , \: t) \: = \: 3 P(\bm{r} \: , \: t) \: = \:
\int_{0}^{\infty} {\rm d} \epsprime \: \epsprime \:
{\cal N}(\bm{r} \: , \: t \: , \: \epsprime)  \; .
\ee
Here $P(\bm{r} \: , \: t)$ is the pressure. Multiplying transport equation
(\ref{simpletransp}) by $\epsprime$ and then integrating over plasma rest-frame
particle energy $\epsprime$ yields, after a partial integration:
\be
\frac{\partial}{\partial x^{\mu}} \left( U^{\mu} \: {\cal W} \right)
+ \: \frac{\left( \nabla \bdot U \right)}{3} \: {\cal W} \: = \: 0 \; .
\ee
This can be written as
\be
\gamma \: \frac{{\rm d} {\cal W}}{{\rm d}t} + \frac{4}{3} \:
\left( \frac{\partial \gamma}{\partial t} + \grad \bdot \bm{U} \right)
\: {\cal W} \: = \: 0 \; ,
\ee
or equivalently
\be
\frac{{\rm d} P}{{\rm d}\tau} \: = \: - \frac{4}{3} \:P
\left( \nabla \bdot U \right) \; ,
\label{eq:dPdt}
\ee
With density relation (\ref{altdens}) this implies that
\be
{\cal W}  \: , \: P \: \propto \: \left( \rhoprime \right)^{4/3} \; ,
\ee
the usual relation for an ideal, adiabatic  and relativistically hot gas.

\subsection{The \NTracer model}
In the simulations presented in the main paper we assume that each jet component
(i.e. jet spine 1, jet sheath 1, \ldots) has its own population of relativistic
particles. Each jet component `$A$' is given a separate tracer
$\theta'_{\rm A}(\bm{r}\:,\:t)$ (also measured in the plasma rest-frame). In the
numerical simulations that are used in this paper, the tracer values of the individual
jet component in a certain grid cell $\bm{r}_i$ at a time $t$ represents the mass
fraction of that jet component (see $\SWa$ or $\SWb$). In other words, the mass
density of jet component $A$ is calculated according to
\mbox{$\rho'_{\rm A}(\bm{r}\:,\:t) =
\theta'_{\rm A}(\bm{r}\:,\:t) \: \rho'(\bm{r}\:,\:t)$}. In particular,
\mbox{$\theta'_{\rm A} = 0$} in a certain grid cell means that jet component $A$
is absent in that grid cell, whereas \mbox{$\theta'_{\rm A} = 1$} means that the
material inside the cell purely consists of jet component $A$. Values in between
\mbox{$0 < \theta'_{\rm A} < 1$} indicate that the grid cell consists of multiple
components. Therefore it is reasonable to assume that the relativistic particle
distribution $\mathcal{N}_{A}$, for a jet component $A$, with mass fraction (or
tracer value) $\theta'_{A}$ in a grid cell $\bm{r}$ at time $t$ can be scaled with
the bulk mass density $\rho$ according to:
\be
	\mathcal{N}_{\rm A} \propto
	\mathcal{N_{\rm A0}}\left(\frac{\rho'_{\rm A}}
   {\rho'_{\rm A0}} \right)^{\frac{s+2}{3}} \;
	= \; \mathcal{N_{\rm A0}}
	\left(\frac{\theta'_{\rm A}\:\rho'}{\rho'_{\rm A0}} \right)^{\frac{s+2}{3}} \; ,
\label{eq:NA-tracer}
\ee
neglecting the effects of synchrotron cooling, and with
$\mathcal{N_{\rm A0}}(\bm{r}\:,\:t)$ and $\rho_{\rm A0}(\bm{r}\:,\:t)$ the number
density and mass density of the relativistic particles of jet component ${\rm A}$
at the injection site (the jet inlet). The total synchrotron emissivity as a
result of all the jet components $A$ is the sum of the synchrotron emissivities
of the individual jet components. In this way, we finally end up with:
\be
j_{\nu} \: = \: \Sigma_{\rm A} j_{\nu, {\rm A}} \:
\propto \Sigma_{\rm A} \: \mathcal{D}^{2+\alpha} \mathcal{N}_{\rm A} \:
(B_{\perp}')^{\alpha + 1} \:(\nu)^{-\alpha} \; .
\ee
Here $\alpha$ is the usual spectral spectral index of optically thin synchrotron
radiation, related to the slope $s$ of the electron distribution by
$\alpha = (s-1)/2$, c.f. \citet{RybickiLightman1986}. 
  
  \subsection{The \NCooling model}
    \label{subsec:TheNCoolingmodel}

In the previous section we calculated the distribution
$\mathcal{N}(\bm{r} \: , \: t)$ of the relativistic electrons for the case that
synchrotron losses can be neglected. This distribution can also be derived for
the case that synchrotron losses are included (for a full derivation, see
\citealt{CamusThesis}, Section 4.3.4). In short the derivation goes as follows:

It is generally believed that electrons are accelerated when crossing (strong)
shocks. Therefore, a shock is assumed to be the injection site of a relativistic
particle population. After crossing the shock, the energy evolution of a single
particle moving along stream lines (of the post-shock plasma) can be written as:

\be
\frac{\rm{d}}{{\rm d}t} \left\{\ln(\epsilon') \right\} \: = \:
\frac{\rm{d}}{{\rm d}t} \left\{ \ln[(n')^{\frac{1}{3}}] \right\} +
\frac{1}{\epsilon'}\left(\frac{\rm{d}\epsilon'_{\rm sy}}{{\rm d}t}\right),
\label{eq:energylosses1}
\ee
the sum of adiabatic losses (the first term on the right hand side) and radiative
energy losses (the second term), and $n'(\bm{r}\:,\:t)$ is
the proper number density of the relativistic leptons, defined as
\be
	n'(\bm{r} \: ,\: t) = \int_{0}^{\infty} {\rm d} \epsilon' \: {\cal N}_{\epsprime}(\bm{r} \: ,\: t) \; .
\ee 
The adiabatic loss term uses that $n'(\bm{r}\:,\:t)$ satisfies (away from shocks)
\be
	\frac{\partial}{\partial x^{\mu}} \left(U^{\mu} \: n'(\bm{r}\:,\:t) \right) = 0 \; ,
\ee
which follows straightforwardly from an integration over comoving energy $\epsilon'$ of Eqn. (\ref{simpletransp}).
This equation, together with Eqn. (\ref{Echange}) for the adiabatic energy loss rate, implies
\be
	\left( \frac{1}{\epsilon'} \: \frac{{\rm d} \epsilon'}{{\rm d} \tau} \right)_{\rm ad} 
	= - \mbox{$\frac{1}{3}$} \left( \grad \bdot U \right)
	= \mbox{$\frac{1}{3}$} \left( \frac{1}{n'} \frac{{\rm d} n'}{{\rm d} \tau} \right) \; . 
\ee
When the relativistic electron moves in a
local FRF magnetic field ${B'}$, the energy loss due to synchrotron emission is:
\be
\left( \frac{{\rm d} \epsilon'}{{\rm d}t} \right)_{\rm sy} \: = \:
-\beta_{\rm sy} (B')^2 (\epsilon')^2 \; ,
\label{eq:synchrotronlosses}
\ee
where $\beta_{\rm sy} \: = \: 2e^4/3m_{\rm e}^2$ with $e$ the unit of electric charge and $m_{\rm e}$ the electron mass, using $c=1$
as before.
\nskip
Substituting \equref{eq:synchrotronlosses} into \equref{eq:energylosses1},
one can rewrite \equref{eq:energylosses1} as:

\be
\frac{\rm{d}}{{\rm d}t}\left\{ \ln\left[\frac{\epsilon'}{(n'_{\rm e})^{1/3}}\right] \right\} \: = \:
-\beta_{\rm sy} (B')^2 (n')^{1/3}\left[\frac{\epsilon'}{(n')^{1/3}}\right] \; .
\label{eq:energylosses2}
\ee
The solution, obtained by time-integration, is
\be
\label{timeint}
	\frac{1}{\tilde{\epsilon'}(t)} - \frac{1}{\tilde{\epsilon'(t_0)}} \: = \:
	\int_{t_{0}}^t\:{\rm d}t'\:\beta_{\rm sy}\:(B')^2\:(n')^{1/3} \: \equiv \: \frac{1}{\tilde{\epsilon'_{\infty}}(t)} \; .
\ee
Here define $\mbox{$\tilde{\epsilon'}(\bm{r}\:,\:t)\: \equiv \:
\epsilon'(\bm{r}\:,\:t)/[n'(\bm{r}\:,\:t)]^{1/3}$}$. Also, $t_0$
corresponds to the time of injection of the particles at the jet inlet.
Relation (\ref{timeint}) defines the quantity $\epsilon'_{\infty}(t) = [n'(t)]^{1/3} \: \tilde{\epsilon'_{\infty}}(t)$, 
the energy where synchrotron losses modify the spectrum.
Note that (by definition) $\epsilon'_{\infty} = \infty$ at $t = t_{0}$.
We can solve for the corresponding energy $\epsilon'(t)$ (for $t \ge t_{0}$):
\be
	\epsilon'(t) = \frac{\displaystyle \epsilon'_{0} \: \epsilon'_{\infty}(t) \:  \left[ n'(t) \right]^{1/3}}
	{\displaystyle  \epsilon'_{0} \: \left[ n'(t) \right]^{1/3} + \epsilon'_{\infty}(t) \: \left[ n'_{0}\right]^{1/3}} \; . 
\label{eq:synloss0}
\ee
Here $n'_{0} = n'(\bm{r}_0\:,\:t_0)$ and $\epsilon'_0 = \epsilon'(\bm{r}_0\:,\:t_0)$. Note that $\epsilon'(t) \le \epsilon'_{0}$, 
as expected, when the jet expands so that $n' \le n'_{0}$.
\nskip
In this work, the relativistic particles are only injected at the jet inlet:
no fresh relativistic particles are injected anywhere along the flow. Since the
jet expands globally as one moves along the jet axis towards the jet-head, there will be
mainly energy losses (except for local turbulence, or near shocks where, locally, 
\mbox{$\grad \bdot \bm{V} < 0$}). 
This determines the energy distribution of the relativistic leptons. Let us define the normalized distribution function
$f(\epsilon'\:,\: \bm{r}\:,\: t)$ by
\be
\label{normdistr}
	f(\epsilon'\:,\: \bm{r}\:,\: t) = \frac{\displaystyle {\cal N}_{\epsprime}(\bm{r} \: ,\: t)}{n'(\bm{r} \: , \: t)} \; \; \Longleftrightarrow \; \; 
	\int_{0}^{\infty} {\rm d} \epsilon'\: f(\epsilon'\:,\: \bm{r}\:,\: t) = 1 \; .
\ee 
We assume $f(\epsilon'\:,\: \bm{r}\:,\: t)$ to satisfy the following initial condition at injection (power-law injection):
\be
\label{injspectr}
	f(\epsilon'_0\:,\:\bm{r}_0\:,\:t_0) =
	\kappa \: (\epsilon'_0)^{-s} \; ,
\label{eq:synlossb}
\ee
with $\kappa$ some constant that ensures the proper normalization of $f(\epsilon'_0\:,\:\bm{r}_0\:,\:t_0)$.
Conservation of the number of particles, thereby preserving the unit normalization of $f(\epsilon'\:,\:\bm{r}\:,\:t)$, leads to:
\be
	f(\epsilon'\:,\:\bm{r}\:,\:t) \: =\:	
	f(\epsilon'_0\:,\:\bm{r}_0\:,\:t_0) \: \left( \frac{{\rm d} \epsilon'}{{\rm d}\epsilon'_0} \right)^{-1} =
	f(\epsilon'_0\:,\:\bm{r}_0\:,\:t_0) \: \left( \frac{\epsilon'}{\epsilon'_0} \right)^{-2} \: \left( \frac{n'}{n'_{0}} \right)^{1/3} \; .
\label{eq:synlossc}
\ee
The second equality follows straightforwardly from relation \equref{eq:synloss0} which yields 
${\rm d} \epsilon' / {\rm d} \epsilon'_{0} = \left( \epsilon' / \epsilon'_{0} \right)^2 \: \left(n'/n'_{0} \right)^{-1/3}$. 
Adopting (\ref{injspectr}) for $f(\epsilon'_0\:,\:\bm{r}_0\:,\:t_0)$
we find:
\be
	f(\epsilon'\:,\:\bm{r}\:,\:t) \: =\: \kappa \left( \epsilon' \right)^{-s} \: \left( \frac{\epsilon'}{\epsilon'_0} \right)^{s-2}
	\: \: \left( \frac{n'}{n'_{0}} \right)^{1/3} \; .
\ee
Definition (\ref{normdistr}) then gives
\be
\label{Nfinal}
	{\cal N}_{\epsprime}(\bm{r} \: ,\: t) = n' \: f(\epsilon'\:,\:\bm{r}\:,\:t) \: =\:
	n'_{0} \: \kappa \left( \epsilon' \right)^{-s} \: \left( \frac{\epsilon'}{\epsilon'_0} \right)^{s-2}
	\:  \left( \frac{n'}{n'_{0}} \right)^{4/3} \; .
\ee
The two factors involving $\epsilon'/\epsilon'_{0}$ and $n'/n'_{0}$ give how much the spectrum is depressed with respect to the original power law 
(as observations show that $s = 2\alpha + 1 > 2$ for most optically thin synchrotron emission) by the combined action of expansion and energy losses.
Using that relation (\ref{eq:synloss0}) can be manipulated into
\be
	\frac{\epsilon'}{\epsilon'_{0}} = \left( \frac{n'}{n'_{0}} \right)^{1/3} \: \left(1 - \frac{\epsilon'}{\epsilon'_{\infty}} \right) \;,
\ee
which incidentally shows that the spectrum must cut off at $\epsilon' = \epsilon'_{\infty}$, one can write ${\cal N}_{\epsprime}(\bm{r} \: ,\: t)$ as
\be
\label{Nfinal2}
	{\cal N}_{\epsprime}(\bm{r} \: ,\: t) \: =\: n'_{0} \: \kappa \left( \epsilon' \right)^{-s} \:
	\left( \frac{n'}{n'_{0}} \right)^{(s+2)/3} \:  \left(1 - \frac{\epsilon'}{\epsilon'_{\infty}} \right)^{s-2} \; .
\ee

\subsection{Implementation in the numerical code}

In order to calculate ${\cal N}_{\epsprime}(\bm{r} \: ,\: t)$, we need to determine the values of
the relativistic proper number density $n'({\bm r}\:,\:t)$; its value at the time of injection $n_{\rm e\:0}'({\bm r}\:,\:t)$; the particle
energy $\epsilon'({\bm r}\:,\:t)$ and the cut-off energy
$\epsilon_{\infty}'({\bm r}\:,\:t)$. In $\SWa$ it is discussed how a number of
hydrodynamical quantities, such as the particle energy $\epsilon'$ are calculated
by the code {\small MPI-AMRVAC}.
By adding three additional transport equations to this system, it is possible to
also calculate the other three quantities
$n'({\bm r}\:,\:t)$, $n'_{\rm e\:0}({\bm r}\:,\:t)$ and
$\epsilon'_{\infty}({\bm r}\:,\:t)$. These additional transport equations, in conservative form,  are:

\be
\nabla_{\mu}\left(n'\:U^{\mu}\right) \: = \: 0 \; ,
\label{eq:TransportEquation_n}
\ee

\be
\nabla_{\mu}\left(n'\:n_{0}'\:U^{\mu}\right) \: = \: 0 \; ,
\label{eq:TransportEquation_n0}
\ee
which is equivalent with $U^{\mu} \nabla_{\mu} n_{0}'=0$ because of Eqn. (\ref{eq:TransportEquation_n}), and
\be
\nabla_{\mu}\left[\epsilon_{\infty}'\:(n')^{2/3}\:U^{\mu}\right] \: = \:
-\tilde{\beta}_{\rm sy}\:(B')^2\:(\epsilon_{\infty}')^2\:(n')^{2/3} \; ,
\label{eq:TransportEquation_epsinf}
\ee
with $\tilde{\beta}_{\rm sy}=2/3\:\beta_{\rm sy}$. For more details see for
example 
\citet{Camus2009} or \citet{CamusThesis}, Eq. 4.102 -- 4.103. 
\nskip
Finally, the synchrotron frequency $\nu'$ of a particle with energy $\epsilon'$,
which moves along a magnetic field $B'$ scales as:

\be
\nu' \propto B'_{\perp} \: (\epsilon')^2 \; ,
\ee
Using this result, we end up with the synchrotron frequency dependent version of
the particle distribution:
\be
\mathcal{N}\: \propto\: \mathcal{N}_0\left(\frac{n'_{\rm e}}{n'_{\rm e\:0}}\right)
^{\frac{s+2}{3}} \left(1 - \sqrt{\frac{\nu'}{\nu'_{\infty}}} \right)^{s-2} \; .
\label{eq:NCooling}
\ee

When we compare this particle distribution to that of the \NTracer
model \equref{eq:N-tracer}, we notice the following:
In case of the \NTracer model, we assume that the population of
relativistic particles is directly proportional to the mass-density of the
proton gas, $n'_{\rm e}(\bm{r} \: , \: t) \propto \rho'(\bm{r} \: , \: t)$ and that
this proportionality remains constant. The exact value of this proportionality
constant does not play a role in the particle distribution, since this constant
is divided out. Therefore, the quantities $\rho'(\bm{r} \: , \: t)$ and
$\rho'_{0}(\bm{r} \: , \: t)$ in \equref{eq:N-tracer} can simply be
replaced by $n'_{\rm e}(\bm{r} \: , \: t)$ and $n'_{\rm e\:0}(\bm{r} \: , \: t)$,
as in \equref{eq:NCooling}. Then, it can be easily seen that for frequencies well
below the cut-off frequency $\nu' << \nu'_{\infty}$ this particle distribution, and
the particle distribution of the \NTracer model \equref{eq:N-tracer}
converge, as they should. However, the \NTracer model allows us to
study the emissivity due to (any combination of) the individual jet components at
low energies, whereas the \NCooling model allows us to study the
emissivity at all frequencies below the cut-off frequency for the full jet.

\section{Magnetic pressure from gas pressure}
\label{sec:MagneticPfromGasP}

The simulations in this paper are hydrodynamical, employing a one-fluid approximation characterized by
a single density $\rho$, pressure $P_{\rm gas}$ and velocity $\bm{V}$.
\footnote{In this section we refrain from writing the apostrophe for quantities measured in the fluid rest frame.
This applies to the pressure $P$, the magnetic field $\bm{B}$ (and related quantities) as well as the proper density $\rho$.} 
The bulk density in a certain grid cell is the sum of all
separate (jet + ambient medium) components, which allows us to consider the bulk gas
pressure $P_{\rm gas}$ as the sum of its partial contributions:

\be
	P_{\rm gas} = P_{\rm am} + P_{\rm jt} \;
   = \; \frac{\rho \mathcal{R} T}{\mu} \;
   = \; \frac{\mathcal{R} T}{\mu} \: \left(\rho_{\rm am} + \rho_{\rm jt}\right) \; .
\ee
Here $\mathcal{R}$ is the gas constant and $\mu$ is the mass of the particles in units
of hydrogen mass. The subscript ``am'' refers to the ambient medium and ``jt''
refers to jet material. In this approximation, we make use of the assumption that all
components within a single grid cell have equal temperature
\mbox{$T \propto P_{\rm gas} / \rho$}. The pressure of the jet
material is taken to be the sum of its individual components:
\be
	P_{\rm jt} = \; \frac{\mathcal{R} T}{\mu} \rho_{\rm jt} \;
          = \; \frac{\rho \mathcal{R} T}{\mu} \: \Sigma_{\rm A} \theta_{\rm A} \;
           = \; P_{\rm gas} \: \Sigma_{\rm A} \theta_{\rm A} \;
           \equiv \Sigma_{\rm A} P_{\rm A} \; ,
\ee
where we used the fact that
$\rho_{\rm jt}=\rho\:\Sigma_{A}\theta_{A} = \rho\:\theta_{\rm jt}$, with
$\theta_{\rm jt}$ being the sum of all jet component mass fractions within a certain
grid cell. Under these assumptions one finds that in the one-fluid approximation,
the contribution of jet component $A$ to the total pressure in a certain grid cell
is equal to
\mbox{$P_{\rm A}(\bm{r},t) = \theta_{\rm A}(\bm{r},t) \: P_{\rm gas}(\bm{r},t)$}.
\nskip
Next we consider the time evolution of the gas pressure. For an ultra-relativistic gas,
the result is already given by Eqn. \equref{eq:dPdt}. In general, the gas pressure satisfies
an equation of state $P \: \propto \: \rho^{\Gamma}$ and satisfies
\be
\frac{{\rm d} P}{{\rm d}\tau} \: = \: - \Gamma \:P
\left( \nabla \bdot U \right) \; ,
\label{eq:PgasEvolution}
\ee
where $\Gamma$ is the adiabatic (or polytropic) index of the (one-fluid
approximation) gas. For a non-relativistic cold gas $\Gamma = 5/3$, and
for an ultra-relativistic gas $\Gamma = 4/3$.
For a gas that is in an intermediate phase, an {\em effective} polytropic index
$\Gamma_{\rm eff}(P \: , \: \rho)$ can be defined. In our simulations we make use of the
Mathews approximation of the Synge equation of state (see for example
\citealt{Synge1957}; \citealt{Blumenthal1976} or \citealt*{Goedbloed2019}
).
In this approximation, the adiabatic index is dependent on the particle energy
and varies between $4/3 < \Gamma_{\rm eff} < 5/3$ (see $\SWa$ for more details).
\halfskip
Now consider a magnetic field \mbox{${\bm B} = {\bm B}({\bm r}\:,\:t)$}, with
field strength \mbox{$B \equiv |{\bm B}|$}, and unit vector
\mbox{$\hat{{\bm b}} = {\bm B}/B$} along the magnetic field.
In the case of ideal MHD, the time evolution of the magnetic pressure
\mbox{$P_{\rm M} = B^2/8\pi$} for a regular flow with relativistic velocity field
${\bm U}(\bm{r}\:,\:t) = \gamma(\bm{r}\:,\:t)  \bm{V}(\bm{r}\:,\:t) $ can be written as (which can easily be seen from
\citealt{Achterberg2018a}, 
Eqn. 6.71):

\be
\frac{{\rm d}P_{\rm M}}{{\rm d}\tau} = \gamma\frac{{\rm d}P_{\rm M}}{{\rm d}t} =
2(\gamma+1)\:P_{\rm M}\:\left[\hat{\bm{b}}\hat{\bm{b}}\: : \:\tilde{\grad}
\left(\frac{\bm{U}}{\gamma+1}\right)
\:- \:\tilde{\grad}\bdot\left(\frac{\bm{U}}{\gamma+1}\right)\right] \; .
\label{eq:PMEvolution}
\ee
The double contraction in this relation is
\be
\hat{{\bm b}}\hat{{\bm b}}\: :
\:\tilde{\grad}\left(\frac{\bm U}{\gamma+1}\right) \: \equiv \:
\left[\left(\hat{{\bm b}}\bdot\tilde{\grad}\right)
\left(\frac{\bm U}{\gamma+1}\right)\right] \bdot\hat{{\bm b}} \; ,
\ee
The operator $\tilde{\grad}$ stands for the fluid rest-frame gradient operator, expressed
via the Lorentz transformation between the fluid rest frame and oberver\rq{s} frame 
in the observer-frame time derivative and spatial derivatives:
\be
\tilde{\grad}\: \equiv \: \grad +
\bm{U}\left(\frac{\partial}{\partial t}\:+\:
\frac{(\bm{U}\bdot\grad)}{\gamma+1}\right) \; .
\ee
In general, \equref{eq:PgasEvolution} and \equref{eq:PMEvolution} do not have the
same form, and a simple relationship between the gas- and magnetic pressure is not available. However,
if the field is isotropically turbulent (entangled) on small scales, the following relation holds:
\be
<b_{i}b_{j}> \: = \: \frac{\delta_{ij}}{3} \; ,
\label{eq:IsotropicallyEntangled}
\ee
Here the average $< \cdot \cdot \cdot >$ is a spatial average over a small volume with a linear size much smaller
than the size of the system. In this case one has
\be
\left< \hat{{\bm b}}\hat{{\bm b}}\: : \:\tilde{\grad}\left(\frac{\bm U}{\gamma+1}\right) \right>
\: \simeq \:
\frac{1}{3}\:\tilde{\grad} \bdot \left(\frac{\bm U}{\gamma+1}\right) \; .
\ee
With this approximation the time evolution of the magnetic pressure collapses to
\be
	\frac{{\rm d}P_{\rm M}}{{\rm d} \tau} \: = \:
	- \mbox{$\frac{4}{3}$} \: P_{\rm M} \: \left[(\gamma+1)\:\tilde{\grad} \bdot
	\left(\frac{\bm{U}}{\gamma+1}\right) \right] = - \mbox{$\frac{4}{3}$} \: P_{\rm M} \: \left( \nabla \bdot U \right) \; ,
\ee
Here we use the relation 
$(\gamma + 1) \tilde{\grad} \bdot \left[ \: \bm{U}/(\gamma+1) \: \right] = \partial \gamma/ \partial t + \grad \bdot \bm{U} = \nabla \bdot U$.
Comparing this with relation (\ref{eq:PgasEvolution}) one sees that the magnetic pressure behaves as an ultra-relativistic gas
with polytropic index $\Gamma_{\rm M} = 4/3$. One concludes that the magnetic pressure scales as
\be
P_{\rm M}'\: \propto\: (\rho')^{4/3}\: \propto\: (n_{\rm e}')^{4/3} \; ,
\ee
where we re-instated the apostrophe notation for quantities in the fluid rest frame.
The ratio of the gas
pressure to the magnetic pressure (the so-called plasma beta) scales as
\be
\frac{P_{\rm gas}'}{P_{\rm M}'} \: \equiv \:
\beta_{\rm p} \: \propto \: (\rho')^{\Gamma - 4/3} \; .
\ee
When the gas is relativistically hot ($\Gamma = 4/3$), the plasma beta
becomes a constant. The energy of the system will be minimized in the case where
$\beta_{\rm p}$ is close to unity. In that case there is equipartition between the
gas pressure and the pressure stored in the magnetic field. This assumption is
often made when trying to get an estimate on the magnetic field strength. However,
in our simulations the plasma beta will still be weakly dependent on the energy
density of the gas through the effective polytropic index $\Gamma$ of the gas.
Finally, we make the assumption that there is no magnetic field present in the
ambient intergalactic medium. In that case, the magnetic pressure and magnetic field strength $B'$ in the
simulations will be completely determined by the jet gas pressure according to:
\be
P_{\rm M}' \: = \: \frac{(B')^2}{8\pi} \:\propto\: P_{\rm jt}'\:
(\rho')^{4/3 - \Gamma_{\rm eff}}\:=\: \left[ \: P_{\rm gas}' \: \Sigma_{\rm A} \theta_{A}' \: \right] \:
(\rho')^{4/3 - \Gamma_{\rm eff}} \; .
\label{eq:B2FromP}
\ee
The quantities $P_{\rm gas}'$, $\theta_{\rm A}'$, $\rho'$ are all calculated by the code, and the effective polytropic index $\Gamma_{\rm eff}$ is
determined at any point in time and space during the simulation. Then
using the assumption of an entangled field will allow us to get a detailed
estimate of the magnetic pressure at any point and time in the flow.

\section{Magnetic field configuration and projection effects}
\label{sec:MagneticField}

As shown in the previous
section, we can approximate the magnetic pressure for an isotropically entangled
magnetic field through the gas pressure, tracer values, mass density and adiabatic
index of the jet components. In the remainder of this section, we expand our
synchrotron emission model further by taking into account the effect of
large-scale (ordered) magnetic fields within the jets. With the ordered fields,
projection effects come into play, as shown in the following sections.

  \subsection{Perpendicular magnetic field component}
\label{subsubsec:PerpendicularMagneticFieldComponent}

As mentioned earlier, the magnitude of the synchrotron emissivity depends on
$B_{\perp}'$, the absolute value of the magnetic field component perpendicular
to the particle's velocity in the plasma rest-frame. 
Therefore, the contribution to the synchrotron emission from relativistic
particles spiralling around a magnetic field $\bm{B}'$ in the
plasma rest-frame, in a direction $\bm{n}'$ (the unit vector
along the line of sight, as measured in the plasma rest-frame) will be determined
by the cross-product of $\bm{B}'$ and $\bm{n}'$, in other words:

\be
B_{\perp}' \; = \; |\bm{B}' \times \bm{n}'| \; = \;
|\bm{B}'| \sqrt{1 - \left(\bm{b}'\bdot\bm{n}'\right)^2} \: ,
\label{eq:Bperpdash}
\ee
where $\bm{b}'$ is the unit vector pointing in the direction of $\bm{B}'$ and
$B' = |\bm{B}'| = \sqrt{\bm{B}' \bdot \bm{B}'}$.
Now consider a photon that is emitted in a direction $\bm{n}'$ in the plasma
rest-frame. When this photon reaches the observer, it appears to
have been emitted from a different direction $\bm{n}$, an effect caused by
general Lorentz transformations on vectors. It can be shown that the relation
between $\bm{n}'$ and $\bm{n}$ is given by
(see for example \citealt{CamusThesis}, Eq. 4.37):

\be
\bm{n'} \; = \; \mathcal{D} \left[ \: \bm{n} \; +\;
\left(\frac{\bm{n}\bdot\bm{\beta}}{V^2}(\gamma - 1) - \gamma \right) \:
\bm{\beta} \right] \; ,
\ee
and this can be rewritten as:
\be
\bm{n'} \; = \; \mathcal{D} \: \bm{n} \; -\;
\left(\mathcal{D}+1\right) \: \frac{\gamma}{\gamma+1} \: \bm{\beta} \; ,
\label{eq:ndash}
\ee
with as before $\bm{\beta}$ the bulk 3-velocity of the plasma in units of $c$;
$\gamma$ the Lorentz factor and $\mathcal{D}$ the Doppler factor. So in a HD
simulation, by assuming a certain magnetic field configuration in the plasma
rest-frame $\bm{B}'$ and choosing a line of sight $\bm{n}$, one can always
calculate the contribution to the synchrotron emissivity from the perpendicular
magnetic field component $B_{\perp}'$ by combining \equref{eq:Bperpdash} and
\equref{eq:ndash}.

In this paper we use 2.5D simulations, where the jets are injected along the
$Z$-direction and the jets are axisymmetrical in the azimuthal ($\phi$-)direction.
Since the simulations are axisymmetrical, the line of sight can simply be chosen
by one free parameter: the viewing angle
$\vartheta = \angle (\hat{\bm{e}}_{Z} ,\: \bm{n})$. The line of sight vector can
in general be written as:

\be
{\bm n} \; = \;
(\sin(\vartheta)\:\cos(\varphi_0),\; \sin(\vartheta)\:\sin(\varphi_0),\;
\cos(\vartheta)) \; .
\ee
The choice for $\varphi_0$ in this case, however, is arbitrary and we take this
to be $\varphi_0 = 0$. Then the line of sight vector simplifies to:

\be
{\bm n} \; = \;
(\sin(\vartheta),\; 0,\; \cos(\vartheta)) \; .
\ee
In the following sections we will discuss the magnetic field configurations that
we consider for our synthesized synchrotron maps. Since we are focussing on
intensity variation maps, the exact scale is arbitrary. This
allows us to set the imposed magnetic field in the plasma rest-frame
$B'$ to order unity at the jet inlet.

 \subsection{Isotropically entangled field configuration}
\label{subsubsec:Bent}

A very entangled magnetic field, on a larger scale does not have a preferable
direction. That means that the magnitude of the perpendicular magnetic field
component $B_{\perp}'$ is not dependent on the viewing angle $\vartheta$, so there
are no projection effects. We find for the perpendicular component of the entangled
field:

\be
(b_{\rm \perp,ent}')^{2} \propto \left|\bm{b}'_{\rm ent} \times \bm{n}'\right|^{2}
\: = \:  (b_{\rm ent}')^{2} \left(1-(\bm{n}'\bdot\bm{b}'_{\rm ent})^{2}\right)
\: = \:  1-(\bm{n}'\bdot\bm{b}'_{\rm ent})^{2} \: .
\label{eq:Bent2}
\ee
Since the field is isotropically entangled, equation
\equref{eq:IsotropicallyEntangled} applies:

\be
\left<(\bm{n}'\bdot\bm{b}'_{\rm ent})^{2}\right> \; = \;
(n')^{i}(n')^{j}\left<b_{i}'b_{j}'\right> \; = \; \frac{1}{3} \; ,
\ee
so that averaging the magnetic field squared on a large enough volume one finds:

\be
\left<(b_{\rm \perp,ent}')^{2}\right> \; = \;
1-\left<\bm{n}'\bdot\bm{b}'_{\rm ent}\right>^{2} \; = \; \frac{2}{3} \; .
\ee
Making use of our earlier result for the magnetic pressure \equref{eq:B2FromP},
we find for the magnitude of the perpendicular field component of the entangled
magnetic field:

\be
(B_{\rm \perp,ent}')^{2}\: \propto\:
\frac{2}{3} \: P_{\rm gas}'\: (\rho')^{\frac{4-3\Gamma}{3}} \:
\Sigma_{A}\theta_{\rm A}'\; .
\label{eq:BentPerp2}
\ee
When the magnetic field configuration is assumed to be purely (isotropically)
entangled, the factor 2/3 can be dropped immediately, since we are merely
studying the intensity variations of the synchrotron emission. However, when
apart from the entangled field, additional ordered field configurations are
assumed, the factor 2/3 needs to be included.

 \subsection{Ordered field configurations}
\label{subsubsec:Bstruc}

In this section we discuss a method to obtain ordered magnetic fields in the
rest frame of the plasma, to calculate their contributions to the
synchrotron emission. The jet is most conveniently described using a cylindrical
coordinate system $(R\:,\:\phi\:,\:Z)$, where the central engine is at rest (the
central engine rest-frame) and the jet propagates outward along the $Z-$axis. In
this reference frame the magnetic field at some point $\bm{r}$ has cylindrical
field components $(B_{R}\:,B_{\phi}\:,B_{Z})$. When the magnetic
field configuration of a jet has a helical structure, the radial field component
$B_{R}$ is zero for a constant jet radius. In that case, the pitch angle
$\kappa$ between the \mbox{azimuthal} field component $B_{\phi}$ and the
poloidal field component $B_{Z}$ can be defined as:

\be
\tan(\kappa) = \frac{B_{Z}}{B_{\phi}} \; ,
\ee
where $\kappa = 0^{\circ}$ corresponds to a pure azimuthal field and
\mbox{$\kappa = 90^{\circ}$} corresponds to a pure poloidal field.
In Cartesian coordinates,
for a jet with a helical magnetic field structure we find:

\be
\bm{B} \; = \;
\left(
  \begin{array}{c}
    - B_{\phi}\sin(\phi) \cos(\kappa) \\
      B_{\phi}\cos(\phi) \cos(\kappa) \\
      B_{Z} \sin(\kappa)
  \end{array}
\right) \; .
\label{eq:Bhelical}
\ee
For a pure poloidal field we write
\mbox{$\bm{B}_{\rm pol} \: = \: B_{\rm pol} \: \bm{b}_{\rm pol}$}, with

\be
\bm{b}_{\rm pol} = \left(
  \begin{array}{c}
    0 \\ 0 \\ 1
  \end{array}
\right) \; .
\label{eq:bpol}
\ee
The magnitude of the poloidal field, $B_{\rm pol}$, is easily determined:
the conservation of poloidal magnetic flux can be written as:

\be
\pi\:R_{\rm jt}^2\:B_{\rm pol}'(R)\; = \; \Phi \; = {\rm constant} \; ,
\ee
where $R_{\rm jt}(Z)$ is the jet radius at a distance $Z$ from the jet inlet.
From this it follows that the magnitude of the poloidal field scales as:

\be
B_{\rm pol}'(R) \; \propto \; \frac{1}{R_{\rm jt}^{2}} \; .
\ee
For a pure azimuthal field we write
\mbox{$\bm{B}_{\rm azm} \: = \: {\rm B}_{\rm azm} \: \bm{b}_{\rm azm}$}, with

\be
\bm{b}_{\rm azm} = \left(
  \begin{array}{c}
    - \sin(\phi) \\ \cos(\phi) \\ 0
  \end{array}
\right) \; .
\label{eq:bazm}
\ee
Also for this magnetic field configuration, the magnitude $B_{\rm azm}$ is
easily determined: Amp\`ere's law determines the azimuthal field in terms of the
current as:

\be
B_{\rm azm}(R)\; = \; \frac{2\:I(<R)}{c\:R} \; ,
\ee
where $I(<R)$ is the current that passes through a disk surface with
radius $R$, centred around the jet axis, and $c$ is the speed of light. For a
homogeneous and structureless jet, a simple assumption can be made that the
electrical current is uniformly distributed over the jet cross section so that
the current density $j$ equals:

\be
j \; = \; \frac{I_{\rm jt}}{\pi\:R_{\rm jt}^2} \; = \; {\rm constant} \; ,
\ee
where $I_{\rm jt}$ is the total electrical current that passes through the jet at
any given point along the jet axis. In that case the electrical current is
proportional to $I(<R) \propto j \: R^2$, so that:

\be
B_{\rm azm}(R) \; \propto \; \frac{R}{R_{\rm jt}^{2}} \; .
\ee

The synchrotron emissivity in a certain direction depends on the field component
perpendicular to the ray direction, all in the local plasma rest frame.
Therefore, we will first transform the magnetic fields back to the plasma 
rest-frame, and after we will calculate the field component perpendicular to
the line of sight.

Magnetic fields transform under Lorentz transformations according to:

\be
\bm{B}' \: = \: \frac{1}{\gamma} \left(
                \bm{B} + \frac{\gamma^2}{\gamma + 1}\bm{\beta} \left(
                \bm{\beta} \cdot \bm{B} \right)
\right) \; ,
\label{eq:Bdash}
\ee
with the meaning $\bm{\beta}$ and $\gamma$ as before.
Now that we have the magnetic field in the plasma rest-frame \equref{eq:Bdash},
as well as the line of sight vector in the plasma rest-frame \equref{eq:ndash},
both in Cartesian coordinates, the magnetic field component perpendicular to
the line of sight is easily obtained:

\be
\bm{B}_{\rm \perp,ord}' \; = \;
\bm{B}'_{\rm ord} \times \bm{n}' \; ,
\label{eq:BordPerp}
\ee
where subscript ``ord" stands for ordered, which can be purely azimuthal,
purely poloidal, or helical. From this field component, we can calculate the
magnetic energy, which is proportional to:

\be
(B_{\rm \perp,ord}')^2 \; = \;
\left|\bm{B}'_{\rm ord} \times \bm{n}'\right|^2 \; = \;
(B_{\rm ord}')^2\left(1-\left({\bm n}'\bdot{\bm b}_{\rm ord}'\right)^2\right) \; .
\label{eq:BordPerp2}
\ee

 \subsection{Mixed field configuration}
\label{subsec:Bmix}

When two components (such as a jet with a spine--sheath structure) start out
with an ordered magnetic field configuration and these components are advected,
turbulence and hydrodynamical instabilities will lead to mixing between these two
components. A jet for example is thought to have a helical magnetic field
structure near the central engine of a AGN. As the jet propagates through the
ambient medium, instabilities and turbulence will lead to a larger fraction of
entanglement. Therefore, it is to be expected that astrophysical jets and their
surrounding cocoon will consist of a mix of large-scale (ordered) fields and
entangled (random) fields. There is another simple argument that predicts the
existence of large-scale, ordered fields: if there are large-scale electrical
currents, such as a current along the jet's axis, they will generate a
large-scale ordered azimuthal field. So even though turbulence is present, if
large-scale currents are not disrupted the large-scale field remains. In the
case where magnetic fields do not play a dynamically important role, one can
make a number of simple assumptions that could describe this evolution based
on pure hydrodynamic simulations.

Consider a system that has an isotropically entangled magnetic field
$\bm{B}_{{\rm ent}}'$ component, as well as an ordered magnetic field component
$\bm{B}_{{\rm ord}}'$ which can be a poloidal field, an azimuthal field or a
helical field. Then the total magnetic field can be written as:

\be
\bm{B}' \; = \; \bm{B}_{{\rm ent}}' + \bm{B}_{{\rm ord}}' \; .
\ee
Since the average over a large enough subvolume yields
$\left<\bm{B}_{{\rm ent}}'\right> = 0$, the average of the total magnetic
pressure is proportional to:

\be
(B')^2 \; \equiv \; \left<(B')^2\right> \; = \;
(B_{{\rm ord}}')^2 + (B_{{\rm ent}}')^2 \; ,
\label{eq:Bmix2}
\ee
where $(B_{{\rm ent}}')^2 = \left<|\bm{B}_{{\rm ent}}'|^2\right>$ is implied.
We can write the magnetic pressure of the field components as a fraction of the
total magnetic pressure in the following way:

\be
\Lambda \; = \;
\frac{(B_{{\rm ent}}')^2}{(B_{{\rm ent}}')^2 + (B_{{\rm ord}}')^2} \; ,
\ee
and

\be
1 - \Lambda \; = \;
\frac{(B_{{\rm ord}}')^2}{(B_{{\rm ent}}')^2 + (B_{{\rm ord}}')^2} \; .
\ee
For the average contribution from the perpendicular magnetic pressure we find:

\be
\left<(B_{\rm \perp}')^2\right> \; = \;
(B_{{\rm \perp,ord}}')^2 + \frac{2}{3}(B_{{\rm ent}}')^2 \; ,
\ee
from which it easily follows that:

\be
(B_{\rm \perp}')^2 \; = \;
(B')^2 \:\left\{\frac{2}{3}\Lambda +
\left(1 - \Lambda\right) \:
   \left|\bm{b}'_{\rm ord} \times \bm{n}'\right|^2
\right\}  \; , 
\ee
or equivalently

\be
(B_{\rm \perp}')^2 \; = \;
(B')^2 \:\left\{\frac{2}{3}\Lambda +
\left(1 - \Lambda\right) \:
   \left(1-(\bm{n}'\bdot\bm{b}'_{\rm ord})^{2}\right)
\right\}  \; ,
\label{eq:BmixPerp2}
\ee
where the average over a large enough subvolume is implied.

This is still a general result. Now if we were to make the assumption that not
just the entangled magnetic field pressure $(B_{\rm ent}')^2$, but the total
magnetic pressure $(B')^2$ scales with the gas pressure as in
\equref{eq:BentPerp2}, the magnetic pressure from the mixed field configuration
\equref{eq:BmixPerp2} can be calculated.
Since $\Lambda$ is the fraction of magnetic pressure from the entangled field
compared to that of the total magnetic field, $\Lambda = 1$ corresponds to
$(B_{\rm ord}')^2 = 0$;
while $\Lambda = 0$ corresponds to $(B_{\rm ent}')^2 = 0$.
For $\Lambda = 0.5$ the magnetic pressure is equally distributed between the ordered
field and the entangled field.

\label{lastpage}

\end{document}